\documentclass[12pt,preprint]{aastex}
\shorttitle{Finding Planet-Search Targets}
\shortauthors{Robinson et al.}
\begin{document}

\title{Identifying Very Metal-Rich Stars with Low-Resolution Spectra:
Finding Planet-Search Targets}

\author{Sarah E. Robinson\altaffilmark{1,3},
Jay Strader\altaffilmark{1,4},
S. Mark Ammons\altaffilmark{1,5},
Gregory Laughlin\altaffilmark{1,6}, and
Debra Fischer\altaffilmark{2,7}}

\altaffiltext{1}{University of California Observatories/Lick
Observatory, Department of Astronomy and Astrophysics, University of
California at Santa Cruz, Interdisciplinary Sciences Building, Santa
Cruz, CA 95064}

\altaffiltext{2}{Department of Physics \& Astronomy, San Francisco State
University, San Francisco, CA 94132}

\altaffiltext{3}{ser@ucolick.org}
\altaffiltext{4}{strader@ucolick.org}
\altaffiltext{5}{ammons@ucolick.org}
\altaffiltext{6}{laughlin@ucolick.org}
\altaffiltext{7}{fischer@stars.sfsu.edu}

\begin{abstract}

We present empirical calibrations that provide estimates of stellar
metallicity, effective temperature and surface gravity as a function
of Lick/IDS indices.  These calibrations have been derived from a
training set of 261 stars for which (1) high-precision measurements
of [Fe/H], $T_{\rm eff}$ and $\log \, g$ have been made using
spectral-synthesis analysis of HIRES spectra, and (2) Lick indices
have also been measured.  Estimation of atmospheric parameters with
low-resolution spectroscopy rather than photometry has the advantage of
producing a highly accurate metallicity calibration, and requires only
one observation per star.  Our calibrations have identified a number
of bright ($V < 9$) metal-rich stars which are now being screened
for hot Jupiter-type planets.  Using the Yonsei-Yale stellar models,
we show that the calibrations provide distance estimates accurate
to $\sim 20\%$ for nearby stars.  We have also investigated the
possibility of constructing a ``planeticity'' calibration, to predict
the presence of planets based on stellar abundance ratios, but find no
evidence that a convincing relation of this type can be established.
High metallicity remains the best single indicator that a given star
is likely to harbor extrasolar planets.


\end{abstract}


\section{Introduction}

In the decade following the announcement of 51 Peg (Mayor \& Queloz
1995) an additional 33 planets with orbital period $P<10 {\rm d}$ have
been discovered.\footnote{See list maintained at www.transitsearch.org}.
This census has revealed a great deal about the properties and evolution
of extrasolar planets.  Short-period planets have relatively high
probabilities of being observed in transit, with eight transiting
planets known as of July 2005.  Among these, HD 209458b (Henry et al.
2000, Charbonneau et al. 2000), TrES-1 (Alonso et al. 2004) and HD
149026b (Sato et al. 2005a) orbit bright parent stars ($V=7.65$,
$V=11.8$ and $V=8.15$, respectively), permitting accurate measurements
of key planetary properties such as mass, radius, albedo and atmospheric
composition.  The models of \cite{bll03} for giant planets without cores
predict that these three planets should have roughly the same radius, so
the observed variation in size is surprising.  \cite{hd149026} invoke a
$70 M_{\oplus}$ solid core for HD 149026b, and \cite{winnholman}
identify obliquity tides as the most viable source of internal heating
to explain the distended nature of HD 209458b.  The radial velocity
surveys have also uncovered unexpected properties of short-period
planets.  The apparent preference for orbits with $P\sim3\, {\rm d}$,
which may imply a mechanism for stopping Type II migration (Lin,
Bodenheimer \& Richardson 1996), is of particular interest.
Additionally, the apparent tidal circularization of orbits with $P<5
{\rm d}$ provides information about the tidal $Q$, giving clues to the
internal structure of planets.

The discovery of more short-period planets, especially hot Jupiters, is
critical to enhancing our understanding of planet formation.  However,
the set of known hot Jupiters is in danger of stagnating:
chromospherically quiet dwarf stars (single or wide-binary members)
brighter than $V = 8$ have, with very few exceptions, been searched for
planets.  Of 27 new planet discoveries reported in 2004, only three have
periods less than ten days, and all of these have $M \sin i <
M_{\rm Saturn}$ (Fischer \& Valenti 2005 and references therein).  The
discovery of increasingly low-mass planets has opened a new line of
inquiry into the differences and similarities between hot Jupiters and
hot Neptunes (Baraffe et al. 2005), but the overall proportion of
short-period planets discovered each year is falling.

In order to find more hot Jupiters, then, one must begin to search
around fainter stars.  There are 1.8 times as many stars with $V > 8$
as with $V < 8$ in the HIPPARCOS catalog alone (Perryman et al. 1997).
Since radial-velocity planet searches are integration-time limited
at any magnitude, faint targets should be chosen with care in order
to make the best use of telescope time.  The best known indicator
of the presence of a short-period planet is metallicity (a partial
list of papers discussing the planet-metallicity correlation is
Gonzalez 1997, Laughlin 2000, Reid 2000, and Fischer \& Valenti 2005).
Therefore, the N2K consortium was created to identify the ``next two
thousand'' metal-rich dwarf stars that would be suitable for
radial-velocity planet searches.

The N2K strategy, described in detail in \cite{n2k1}, consists of a
series of metallicity screenings of increasing precision on late-type
dwarf stars with $V < 11$.  Potential metal-rich stars are first
identified on the basis of broadband photometric models (Ammons et al.
2005, in preparation) with $\sigma_{\rm [Fe/H]} = 0.15$ dex for $V <
9$.  This paper is concerned with the second part of the screening
process, where the metal-rich nature of the candidate stars is confirmed
with low-resolution spectroscopy.  Stars with confirmed super-Solar
metallicity are promoted to a quick-look program of four high-precision
radial velocity observations at large telescopes including Keck, Subaru
and Magellan.  Stars emerging from the quick-look program with RV RMS
$\geq 2\sigma$ then receive follow-up observations to check for
hot-Jupiter-type companions (Fischer et al. 2005).  Finally, the
detected hot Jupiters are subjected to a photometric search for
transits.  The first transiting planet to energe from this strategy is
HD 149026b, whose small photometric depth would render it difficult to
detect through large-scale surveys (e.g. Horne 2002).

This work reports high-precision fits of [Fe/H], $T_{\rm eff}$ and $\log
\, g$ as a function of Lick indices for FGK dwarfs.  Fits between
Str\"{o}mgren indices and [Fe/H] by \cite{SN89} and \cite{sarah} have
been successfully used to select targets for planet searches.  However,
the \cite{uvby} $uvby$ database has already been mined (see also
N\"{o}rdstrom et al. 2004), which spurred us to develop a new mode of
surveying large numbers of stars.  The extended Lick/IDS system (Trager
et al. 1998) comprises broad spectral features between 4000 and 6000
\AA\ that are highly sensitive to stellar atmospheric parameters (see
Gorgas et al. 1993 and Worthey et al. 1994, hereafter W94, for empirical
fits of Lick indices as a function of $T_{\rm eff}$, [Fe/H] and $\log \,
g$; see Korn, Maraston \& Thomas 2005 for an updated discussion of Lick
indices and the physics of stellar atmospheres).  Our calibration has
precision $\pm 0.07$ dex, on par with [Fe/H] measurements from most
high-resolution data.


Metal-rich stars identified with these calibrations can also be used
to study stellar and Galactic evolution in the Solar neighborhood.
A survey of such stars from the Hipparcos catalog (Robinson et
al. 2005, in preparation) is currently in progress.  Additionally, we
can use the stars' atmospheric parameters in conjunction with stellar
models to estimate their absolute magnitude and distance from the Sun.
Since nearby, metal-rich stars should be chemically similar to the Sun,
we hope to enable further understanding of the formation and evolution
of Sun-like stars and the subset of those stars that harbor planets.

\section{Observations}

In order to construct fits for $T_{\rm eff}$, [Fe/H] and $\log \, g$ as
a function of Lick indices, we needed a training set of stars with known
Lick indices and atmospheric parameters.  \cite{sme}, hereafter SME,
present highly precise atmospheric parameters ($\sigma_{[\rm Fe/H]} =
0.03$ dex, $\sigma_{T_{\rm eff}} = 44$K, and $\sigma_{\log \, g} =
0.06$ dex) derived from analysis of high-resolution spectra as
part of the Keck/Lick/AAT planet search.  Because of its precision and
its uniform properties (all observations were taken by the same
observer, with the same instrument, and processed with the same
software), this data set is an ideal benchmark for any project with the
goal of measuring stellar atmospheric parameters.  We obtained 307
low-resolution spectra of 261 stars in the VF05 catalog as a training
set for our calibrations.

Our observations were taken at two telescopes, the Nickel 1m at Lick
Observatory and the 2.1m at Kitt Peak National Observatory.  The Nickel
observations were taken during 2004 April 23-26 and 2004 July 13-15.  We
used the Nickel Spectrograph with a 600 lines/mm grism with spectral
coverage 4100-6000 \AA\ and inverse resolution $R = \lambda / \Delta
\lambda = 500$ (FWHM = 9.6 \AA) at 4800 \AA.  These spectra have $S/N$
$\sim$150 per resolution element (50 per \AA) in the Ca4227 index, the
shortest-wavelength line we measured, increasing to $\sim$ 300 (100 per
\AA) in the Na D line.  The 2.1m observations were taken during 2004
August 27--September 2 with the GoldCam spectrograph, using a 600
lines/mm grism blazed at 4900 \AA.  The spectral coverage was 3800-6200
\AA\ with $R = 1360$ (FWHM = 3.7 \AA) at 5000 \AA.  The spectra have
$S/N$ $\sim$ 230 per resolution element (120 per \AA) in the Ca4227 line
and $\sim$ 380 (200 per \AA) in the Na D index.  As Lick indices are
independent of absolute flux levels (Worthey \& Ottaviani 1997), our
spectra were not flux-calibrated.

\subsection{Measuring Lick Indices}

We measured indices of atomic and molecular lines in our data using the
Lick/IDS system as defined by \cite{FBD77}, extended by W94, and
extended once more by \cite{WO97} to include H$\gamma$ and H$\delta$.
The W94 bandpass definitions were updated by \cite{trager}; it is this
set of bandpasses plus the H$\gamma_F$ index (from Worthey \& Ottaviani
1997) that we use in our analysis.  The atomic line indices are measured
by calculating equivalent widths relative to a pseudocontinuum
interpolated from bandpasses on either side of the spectral line:
\begin{equation} EW = \int  \left ( \left [ f_c(\lambda) - f(\lambda)
\right ] / f_c(\lambda) \right ) d\lambda = \Delta_f  \left ( 1 -
\langle f / f_c \rangle \right ) \label{ew} \end{equation} Molecular
line indices are measured on a magnitude scale, such that
\begin{equation} I_{MAG} = -2.5 \log \left [ \int \left [ f(\lambda) /
f_c(\lambda) \right ] d\lambda / \Delta \lambda_f \right ] = -2.5 \log
\langle f / f_c \rangle \label{mag} \end{equation} where $\Delta
\lambda_f$ is the width of the bandpass centered on the absorption line
and $f_c$ is the pseudocontinuum flux.  We measured Lick indices in our
spectra using the publicly available \texttt{indexf} code by Cardiel,
Gorgas \& Cenarro (\copyright July 11, 2002), which incorporates the
error analysis techniques of \cite{indexf}.  Since the resolution of the
spectra obtained with the Nickel telescope was slightly lower than the
original IDS spectral resolution in most regions of the spectrum, we
could not match the IDS resolution, which slightly increases the index
error (Worthey \& Ottaviani 1997).

\begin{figure}
\plotone{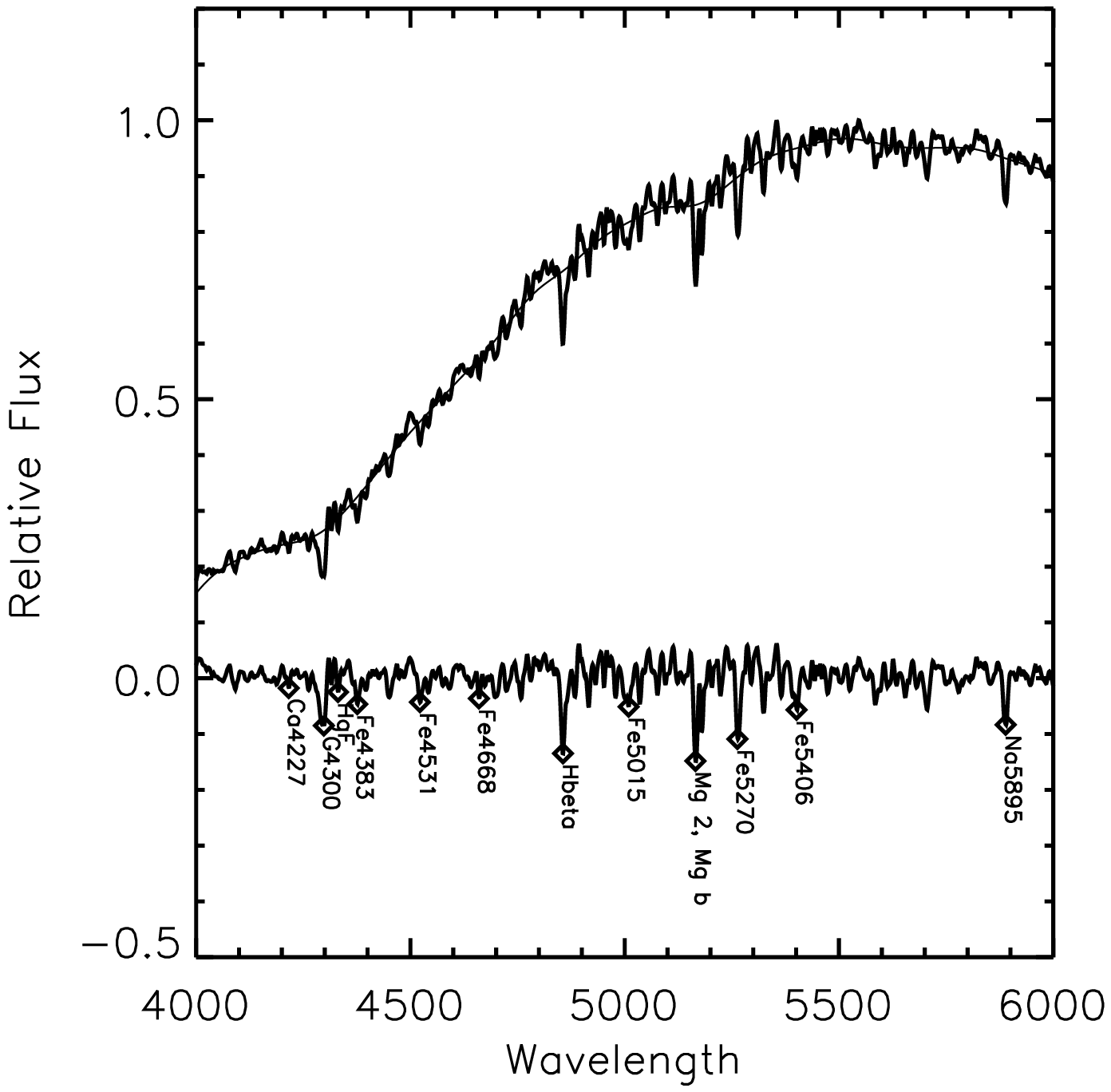}
\caption{Unsharp masking method.  A smoothed version of each spectrum
is subtracted from the original spectrum.  Line centers are then
located in the unsharp-masked spectrum by searching for minima in a
$20 \: \AA$ window around where each line should fall.  This spectrum
of HD 117176 was taken with the Nickel spectrograph.}
\label{findlines}
\end{figure}

Because of flexure in the Nickel CCD spectrograph, we were able to
obtain only rough wavelength solutions that were in general accurate to
$\sigma \sim 10$ \AA.  The GCAM wavelength solutions were accurate to
$\sigma \sim 4$ \AA.  It was therefore necessary to recenter each
spectral line before measuring Lick indices.  This consideration
prompted us to drop the indices CN$_1$, CN$_2$, Mg$_1$, TiO$_1$ and
TiO$_2$ from our analysis: these have multiple spectral lines in the
same central bandpass, making the line center difficult to pinpoint.  To
locate line centers in our data, we used an unsharp masking algorithm,
smoothing each spectrum with a Gaussian kernel of FWHM = 141 \AA\ and
subtracting the smoothed spectrum from the original spectrum.  We then
searched the unsharp-masked spectra for local minima within 20 \AA\ of
each known line wavelength (Figure \ref{findlines}).  Comparing line
centers found by the automatic recentering program with those measured
by hand using Gaussian-fit tools in IRAF for three spectra led us to
estimate an error of $\pm 2$ \AA\ in our recentered wavelength
solutions.  According to W94, the contribution of wavelength errors of
this magnitude to errors in measuring Lick indices is negligible.

\subsection{Matching the Lick/IDS System}

\begin{figure}
\plotone{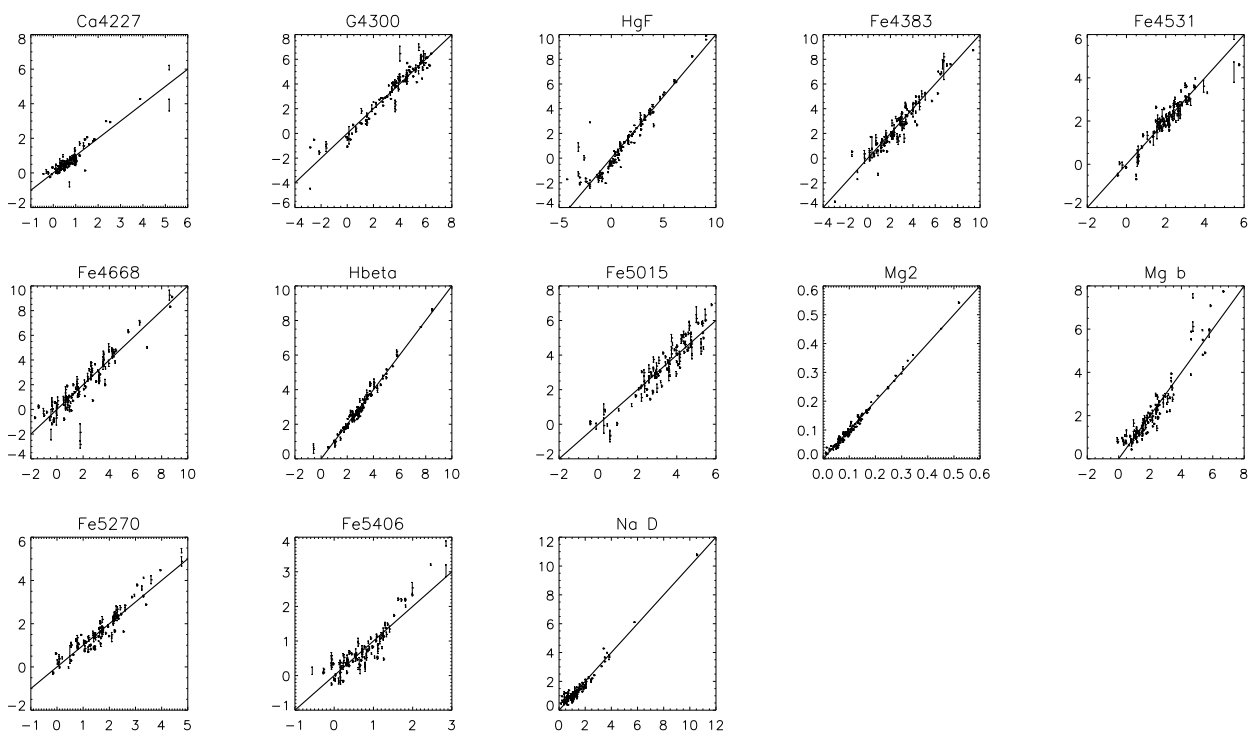}
\caption{Comparison between observed equivalent widths (y-axis) and
those published in \cite{W94} (x-axis) for the 12 indices used in
our fits to stellar atmospheric parameters.  Strongly deviant points
in metal-line measurements correspond to metal-poor stars ([Fe/H]
$\le -1.5$ dex); deviant points in Balmer-line measurements correspond
to cool stars (spectral type K4 or later).  These points were not
used to transform our data to the Lick/IDS system.}
\label{lickmatch}
\end{figure}

We transformed our data to the Lick/IDS system using observations of
Lick/IDS standard stars, which have indices reported in W94 and
\cite{WO97}\footnote{see list maintained at
astro.wsu.edu/ftp/WO97/export.dat}.  48 observations of 29 stars were
taken at the Nickel telescope, and 79 observations of 62 stars were
taken at the 2.1m telescope.  The observed set of Lick standard stars
was chosen to be as similar as possible to our program stars: mainly FGK
dwarfs, with a few BA dwarfs to fill in parts of the sky where FGK stars
were not available.  For each index, we used least-squares analysis to
find a linear fit between the equivalent width published in W94 and that
in our data, creating separate fits for the Nickel and 2.1m data.  In
order that the fits for the metallic indices would not be biased by
extremely metal-poor stars, which show very weak Fe, Ca and Mg lines,
data points that were more than 3 standard deviations from the line of
best fit were rejected and the fits were computed again.  Rejecting
deviant points was also useful for computing fits for the indices
measuring Balmer lines, H$\gamma_F$ and H$\beta$, since our sample
included late K stars with no discernible Balmer absorption.
Encouragingly, the slopes of the linear fits were near unity for almost
all indices.  Notable exceptions were the Ca4455 and Fe5335 lines, which
are highly sensitive to small wavelength shifts; these were excluded
from further analysis.  Also excluded from our fits were Fe5709 and
Fe5782, weak lines that give a small range of possible index
values---our measurements of these were somewhat scattered around those
published in W94.  We retained a sample of 13 indices.  Transformations
from observed indices to values matching W94 are given in Table
\ref{transformations}, along with the error of each index.  Figure
\ref{lickmatch} shows the comparison between our observations and the
published index values for the Lick/IDS calibrator stars in our sample.
Table \ref{starindices} gives the final set of measured Lick indices for
all stars in our training set.
\notetoeditor{We would like Tables 1 and 2 to appear in the electronic
edition only}

\section{Fits to Atmospheric Parameters}

A few previous studies have used Lick indices to determine or confirm
fiducial parameters of individual stars or clusters.  \cite{gorgas93}
created a set of empirical polynomials giving each of the original Lick
indices (Faber, Burstein \& Dresslen 1977) as functions of $\Theta =
5040 / T_{\rm eff}$, [Fe/H] and $\log \, g$; W94 refined these fits and
added polynomials for the 10 indices they added to the system.
\cite{BS94} created a similar set of fitting functions for the Fe5270
and H$\beta$ indices.  \cite{lickfuncs}, hereafter WJ03, inverted the
fitting functions of W94 to find the metal abundances of NGC 188 and NGC
6791.  \cite{buzzoni} measured Lick indices of 139 stars and confirmed
the atmospheric parameters for 91 stars calculated by \cite{bsmr} by
comparing the observed Lick indices with the values calculated by using
the \cite{BS94} fitting functions.

We take a different approach to using Lick indices from previous
authors---since the main goal of this project is to streamline planet
searches, the accurate determination of [Fe/H] is our top priority,
rather than the characterization of how each Lick index behaves as a
function of atmospheric parameters.  Additionally, we are using our
fits to calculate the atmospheric parameters, particularly [Fe/H],
for single, field stars, not cluster members or integrated-light
populations.  We therefore are not able to combine measurements
for many stars or pointings to calculate a mean value of [Fe/H]
for our targets.  For these reasons, we chose to create calibrations
that give $T_{\rm eff}$, [Fe/H] and $\log \, g$ as functions of Lick
indices, rather than defining fitting functions analogous to those of
\cite{BS94}, \cite{gorgas93} or W94.  Our approach to finding stellar
atmospheric parameters follows that of \cite{SN89} and \cite{sarah},
using Lick indices as the independent variables instead of narrow-band
photometric indices.

\S \ref{fitmethods} gives the properties of the stars used to create
our fits and the details of our fitting methods, \S \ref{error}
describes our method of error analysis, and \S \ref{teff}, \ref{feh}
\& \ref{logg} report the fits for $T_{\rm eff}$, [Fe/H] and $\log \,
g$, respectively.  In \S \ref{litcompare}, we compare our work with
that of similar, previously published studies.

\subsection{Training Set and Fitting Methods}
\label{fitmethods}

We began by using the Levenberg-Marquardt method (see, e.g., Press et
al. 1992), to optimize the coefficients, $A_n$, of a linear fit in Lick
indices so as to best reproduce the VF05 atmospheric-parameter values.
Since $T_{\rm eff}$ is the most easily measurable of the atmospheric
parameters (accessible to broadband photometry, with scales between
different investigators matching well), we also tested if more refined
fits were possible for [Fe/H] and $\log \, g$ by adding $T_{\rm eff}$
as term $n+1$.  Since the Levenberg-Marquardt algorithm provides
local convergence around a series of initial guesses for the fit
coefficients, we tested several variations of the initial guesses.
The fits to $T_{\rm eff}$ and $\log \, g$ converged to identical
optimized coefficients each time.  The most refined [Fe/H] calibration
was obtained by first calculating a set of coefficients to the Lick
indices without an additive constant, then using these coefficients
as the initial guesses for a fit that included a constant.  Although
Fe5406 was initially included in the set of indices used to build the
fits, it was always one of the least significant terms and was excluded
from the final versions of each calibration.  Our calibrations apply
to FGK dwarfs, $-1.4 < {\rm [Fe/H]} < 0.54$ dex, $3910 < T_{\rm eff} < 6390$K, and $3.5 < \log \, g < 5.2$ dex.

\subsection{Error Analysis}
\label{error}

We used a variant of the two-phase cross-validation method (Weiss \&
Kulikowski 1991) to determine the uncertainty and accuracy of all
calibrations presented here.  Our error-analysis procedure was as
follows:
\begin{enumerate}
\item The measurements of Lick indices for VF05 stars were randomly
divided into two subsets, \texttt{a} (154 observations) and \texttt{b}
(153 observations).
\item Coefficients of fits for $T_{\rm eff}$, [Fe/H] and $\log \,
g$ as a function of Lick indices were calculated using only subset
\texttt{a} as the training set.
\item To test their accuracy, these fits were used to calculate
atmospheric parameters for the stars in set \texttt{b}.  We found
the fit residuals as, for example, \[ \Delta {\rm [Fe/H] = (Calculated\ 
[Fe/H]) - (VF05\ [Fe/H])}. \]
\item The two sets of stars were swapped: We used subset \texttt{b} 
as a training set to find slightly different versions of the fits
between Lick indices and atmospheric parameters, and calculated
atmospheric parameters and fit residuals using subset \texttt{a}.
\item The residuals from the separate tests on subsets \texttt{a} and
\texttt{b} were combined to perform error analysis.
\end{enumerate}
This method enabled a fully independent verification of the performance
of the calibration method on stars that were not used to determine the
fit coefficients, and should be the most rigorous possible test of our
calibrations.  In order to sample the parameter space of Lick indices
as finely as possible, the published coefficients were calculated
using all VF05 stars with measured Lick indices for the training set.  
Therefore, the true uncertainty of each fit should be even smaller than
what we report.

The two-phase cross-validation showed that the mean and standard
deviation of our calibrations change negligibly when the fits are
calculated using different subsets of the training set.  This means
that the fits are heavily overdetermined, which is essential for
numerical stability.  The uncertainty of each calibration was measured
by fitting a Gaussian to a histogram of the test-set residuals and
measuring the Gaussian FWHM, $\sigma = {\rm FWHM}/2.35$.  Accuracy was
verified by measuring the displacement from zero of the center of
the Gaussian distribution.

\subsection{$T_{\rm eff}$ Calibration}
\label{teff}

\begin{figure}
\plottwo{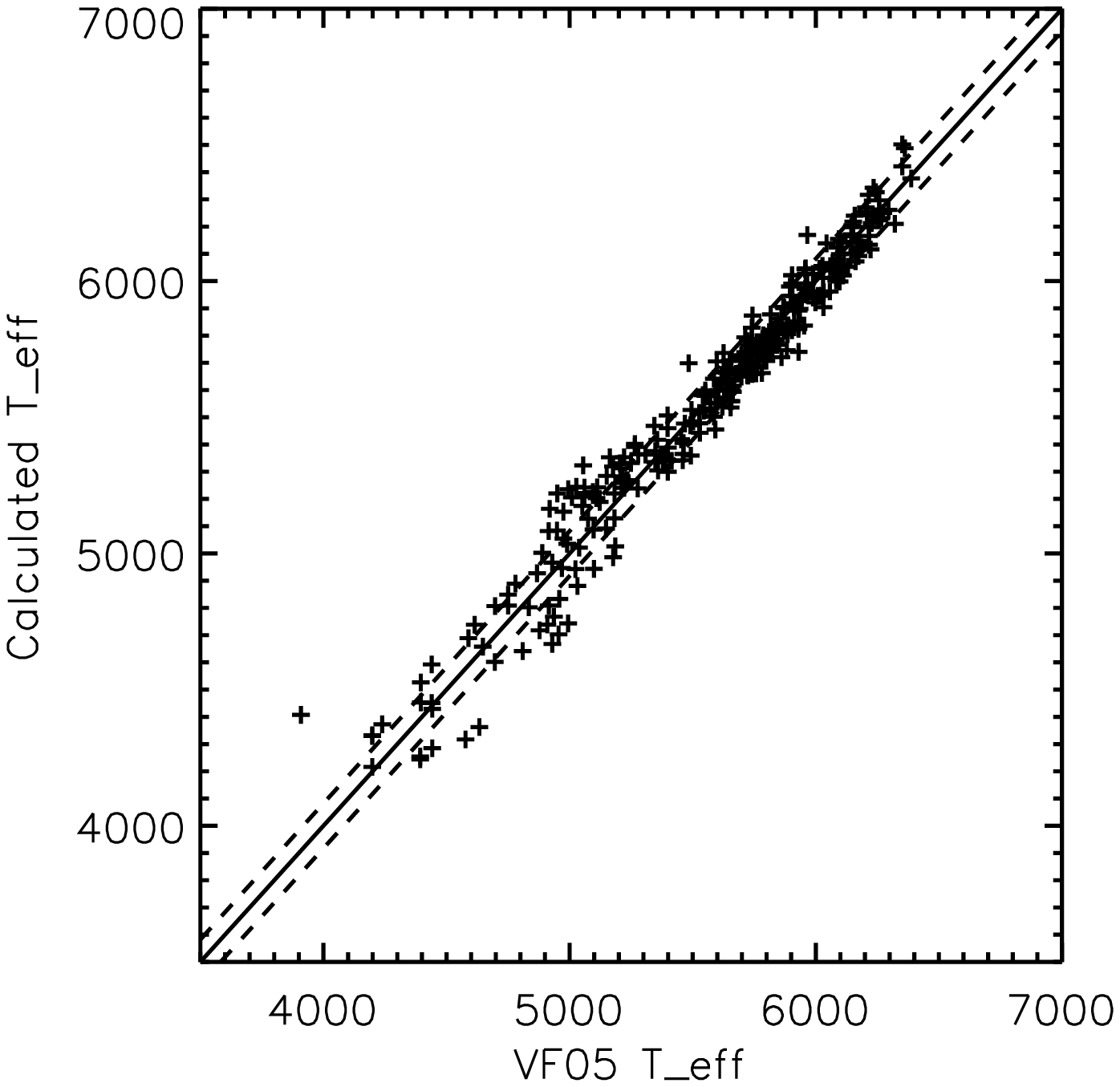}{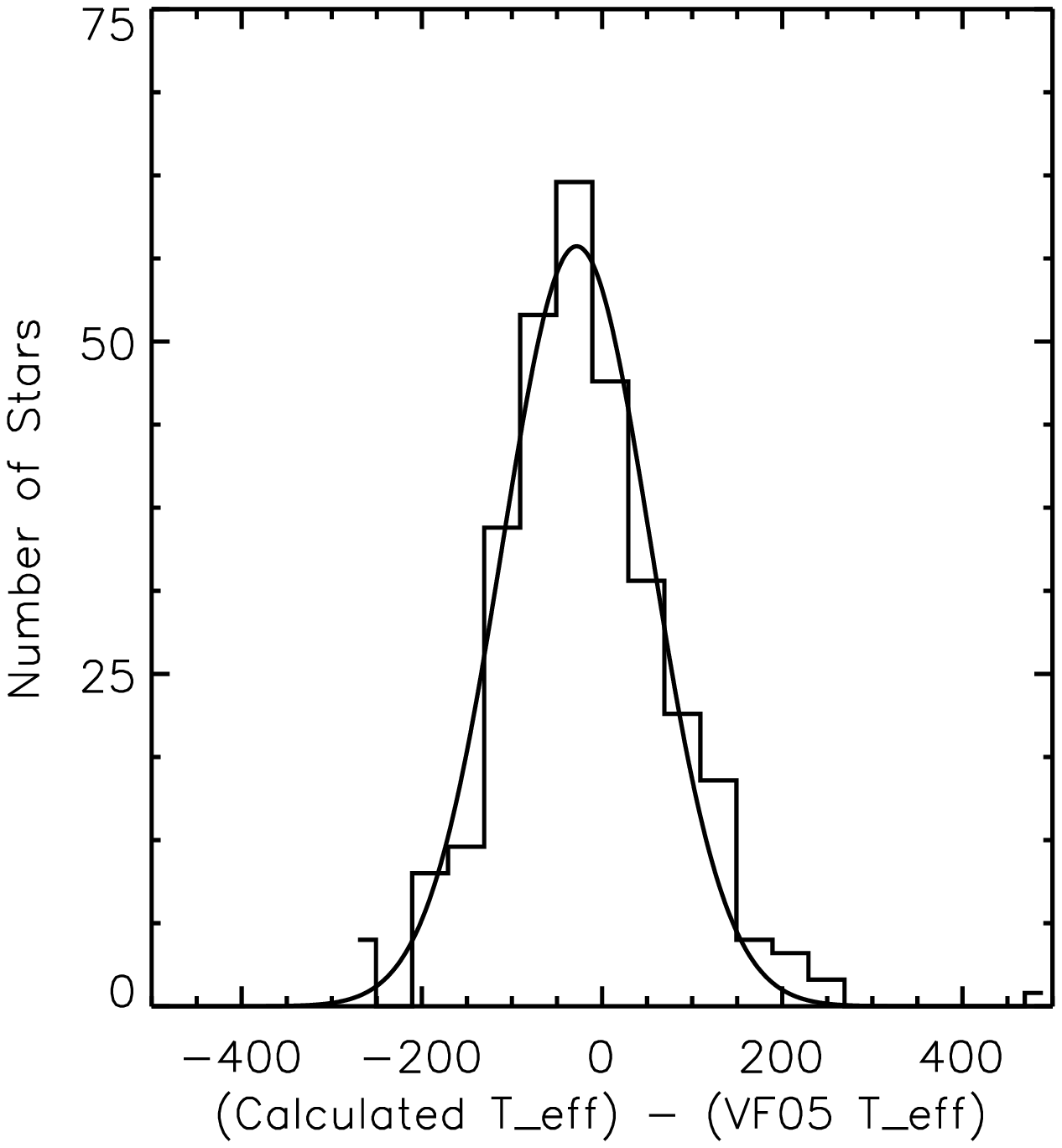}
\caption{Left: Scatter plot showing the performance of the $T_{\rm
eff}$ calibration.  The solid line shows a theoretical perfect
fit, and the dotted lines show the calibration's 1-$\sigma$ error.
Right: Histogram of the residuals of the $T_{\rm eff}$ calibration.
The residuals are well modeled by a Gaussian distribution with $\sigma
= 82$K, centered at -28.4K.}
\label{tefffig}
\end{figure}

The $T_{\rm eff}$ calibration was produced by a linear fit to the Lick
indices in our sample.  It has a reduced $\chi^2$ statistic of 4.52 and
an uncertainty $\sigma_{ T_{\rm eff}} = 82$ K, not highly in excess
of the uncertainties in the VF05 dataset.  The coefficients,
uncertainty and useful range of the $T_{\rm eff}$ calibration are
presented in Table \ref{calibrations} and the performance of this fit
in replicating the atmospheric parameters of the test set is shown
in Figure \ref{tefffig}.  The bump around 5000K, where temperatures
are slightly overestimated, corresponds to the disappearance of
the Balmer lines, which are highly significant in our calibration.
However, since other lines (notably the magnesium features) are strong
temperature indicators for stars cooler than 5000K, the calibration
still functions below this limit.

\subsection{[Fe/H] Calibration}
\label{feh}

\begin{figure}
\plottwo{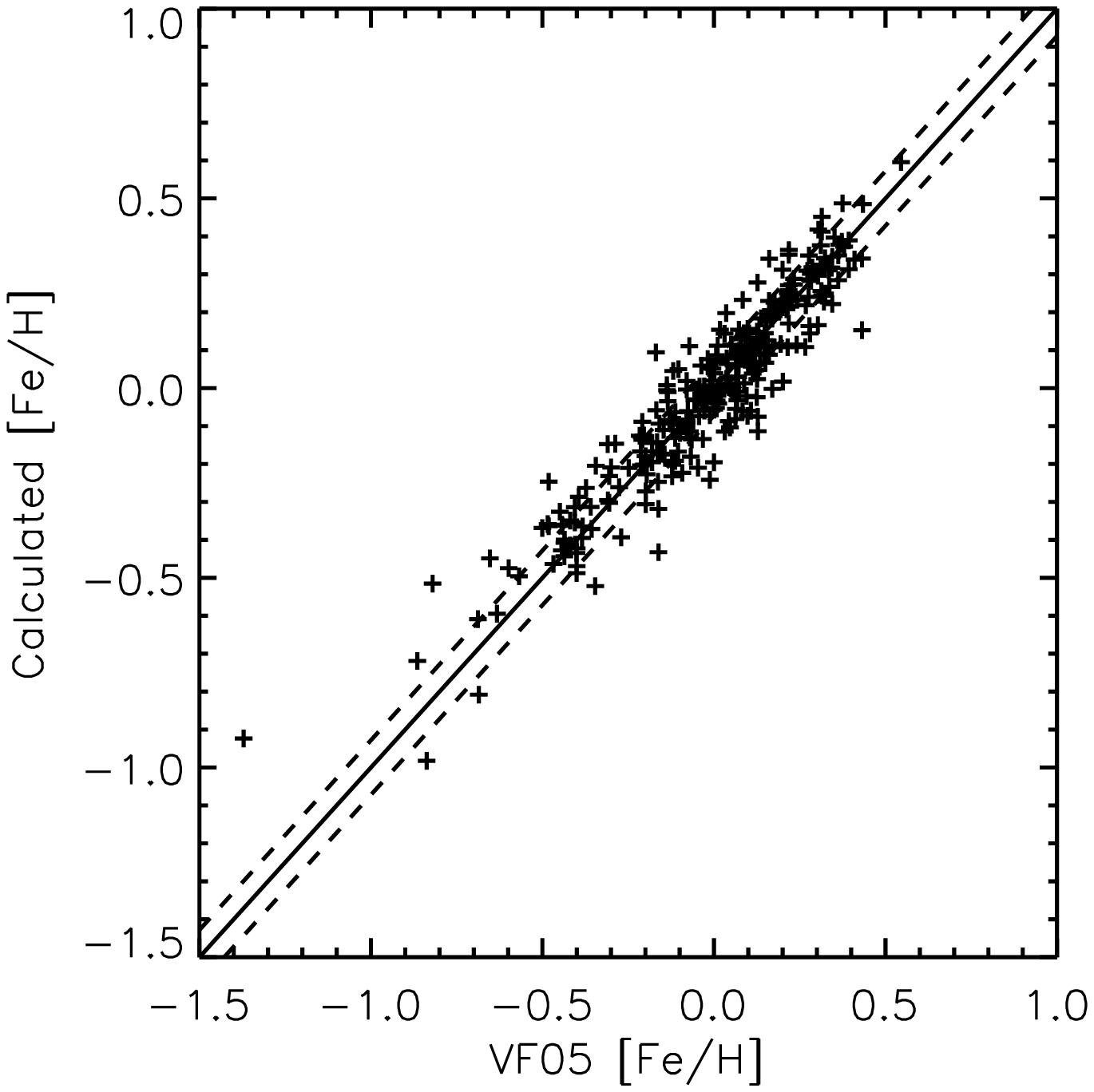}{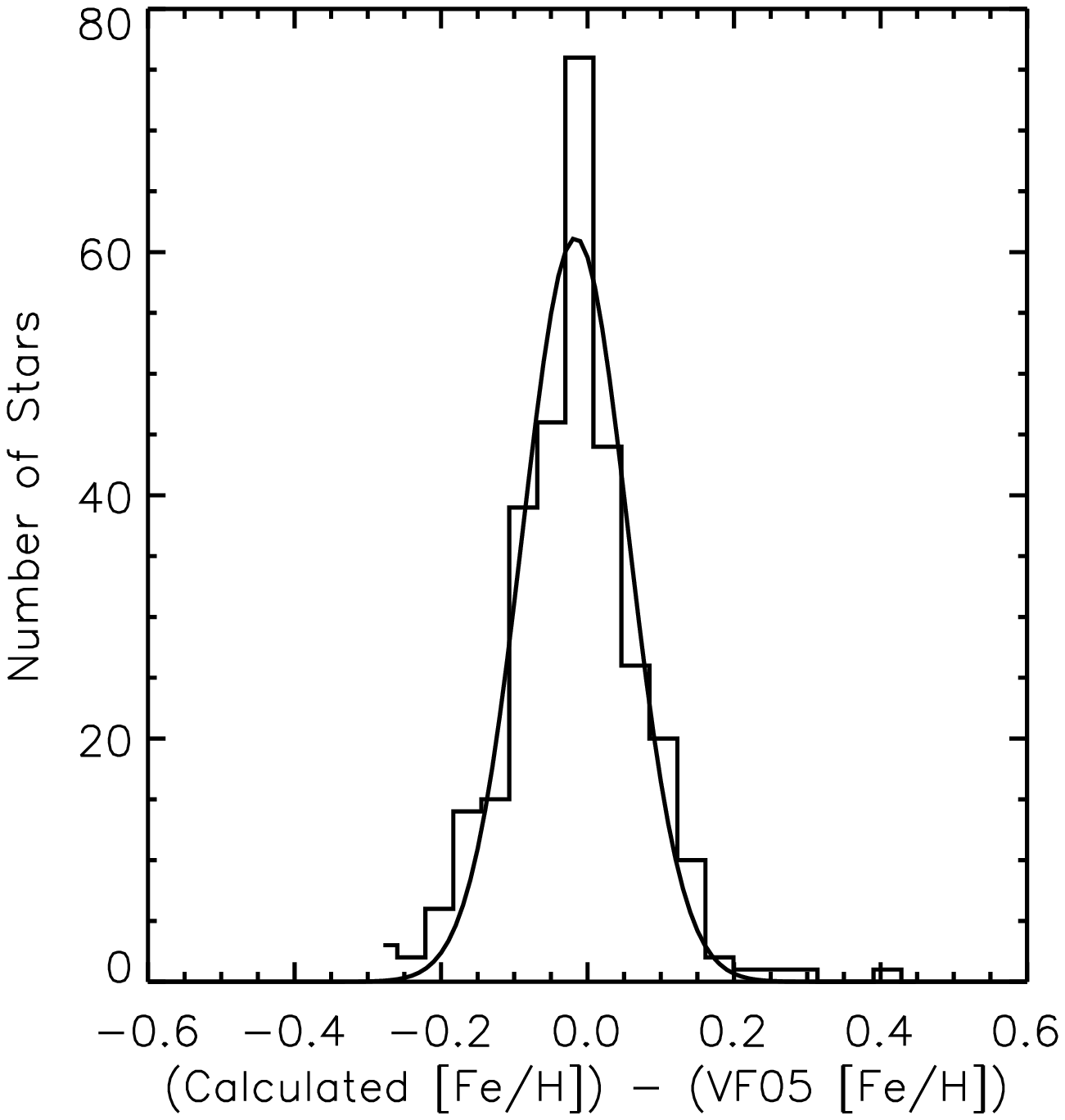}
\caption{Left: Scatter plot showing the performance of the [Fe/H]
calibration.  The solid line shows a theoretical perfect fit, and the
dotted lines show the calibration's 1-$\sigma$ error.  Right: Histogram
of the residuals of the [Fe/H] calibration.  The residuals are
well modeled by a Gaussian distribution with $\sigma = 0.07$ dex,
centered at -0.017 dex.}
\label{fehfig}
\end{figure}

Calibrations based solely on the Lick indices proved not to be the most
robust way to measure [Fe/H] and $\log \, g$.  Line strengths in our
stars should be determined principally by effective temperature, since
all are cool dwarfs that have measurable iron content.  We therefore
conjectured that using $T_{\rm eff}$ as a parameter in the [Fe/H]
calibration would improve the fit.  The resulting calibration has
uncertainty $\sigma_{\rm [Fe/H]} = 0.07$\ dex and a reduced $\chi^2$
statistic of 6.75.  The coefficients and relevant statistics of the
calibration are given in Table \ref{calibrations} and the fit scatter
plot and histogram are shown in Figure \ref{fehfig}.

The metallicities of the test-set stars, which we used to measure
the fit uncertainty, were computed using values of $T_{\rm eff}$
calculated by our $T_{\rm eff}$ calibration.  It should be noted that
[Fe/H] measurements taken by this method will actually be a single
linear combination of Lick indices resulting from the sequential
application of the $T_{\rm eff}$ and [Fe/H] calibrations.  We choose
to leave the fit in terms of $T_{\rm eff}$, emphasizing that the
temperature information of VF05 was necessary to create this fit,
and leaving open the possibility of combining the [Fe/H] calibration
with other ways of measuring $T_{\rm eff}$.

According to \cite{FV05}, an increase in stellar metallicity of 0.2
dex corresponds to a fivefold increase in the probability of planet
detection.  The steepness of this correlation means that it is vital
for the efficiency of planet searches to have as precise [Fe/H]
measurements as possible before proceeding to large telescopes.
If our calibration measures [Fe/H] = 0.20 dex, the probability of
finding a planet is only reduced by $\sim 35\%$ if the star's actual
metallicity is $\sigma_{\rm [Fe/H]} = 0.07$ dex lower than than
that reported value.  This calibration is therefore able to provide
extremely secure target lists for radial-velocity planet searches.

\subsection{$\log \, g$\ Calibration}
\label{logg}

\begin{figure}
\plottwo{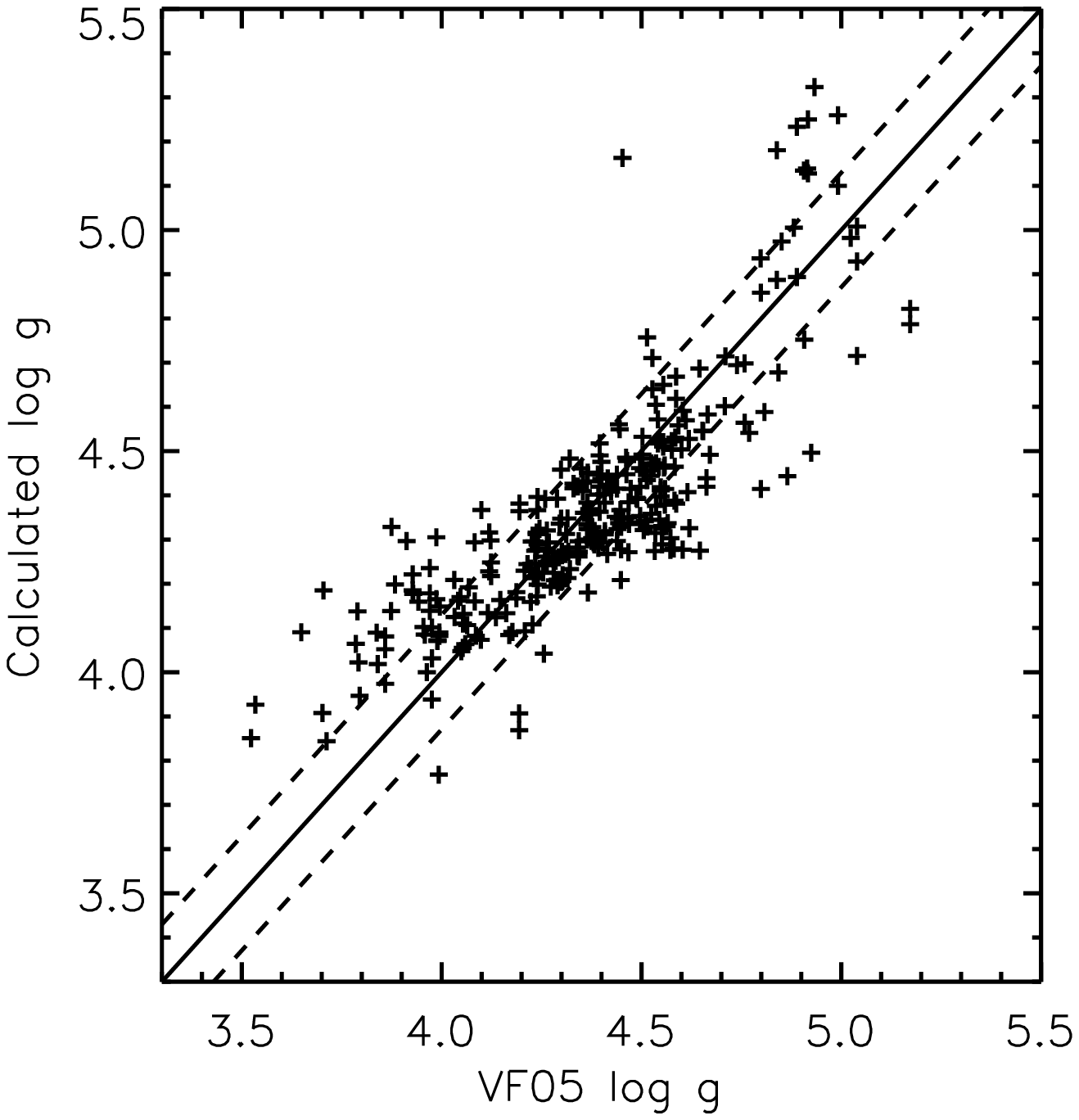}{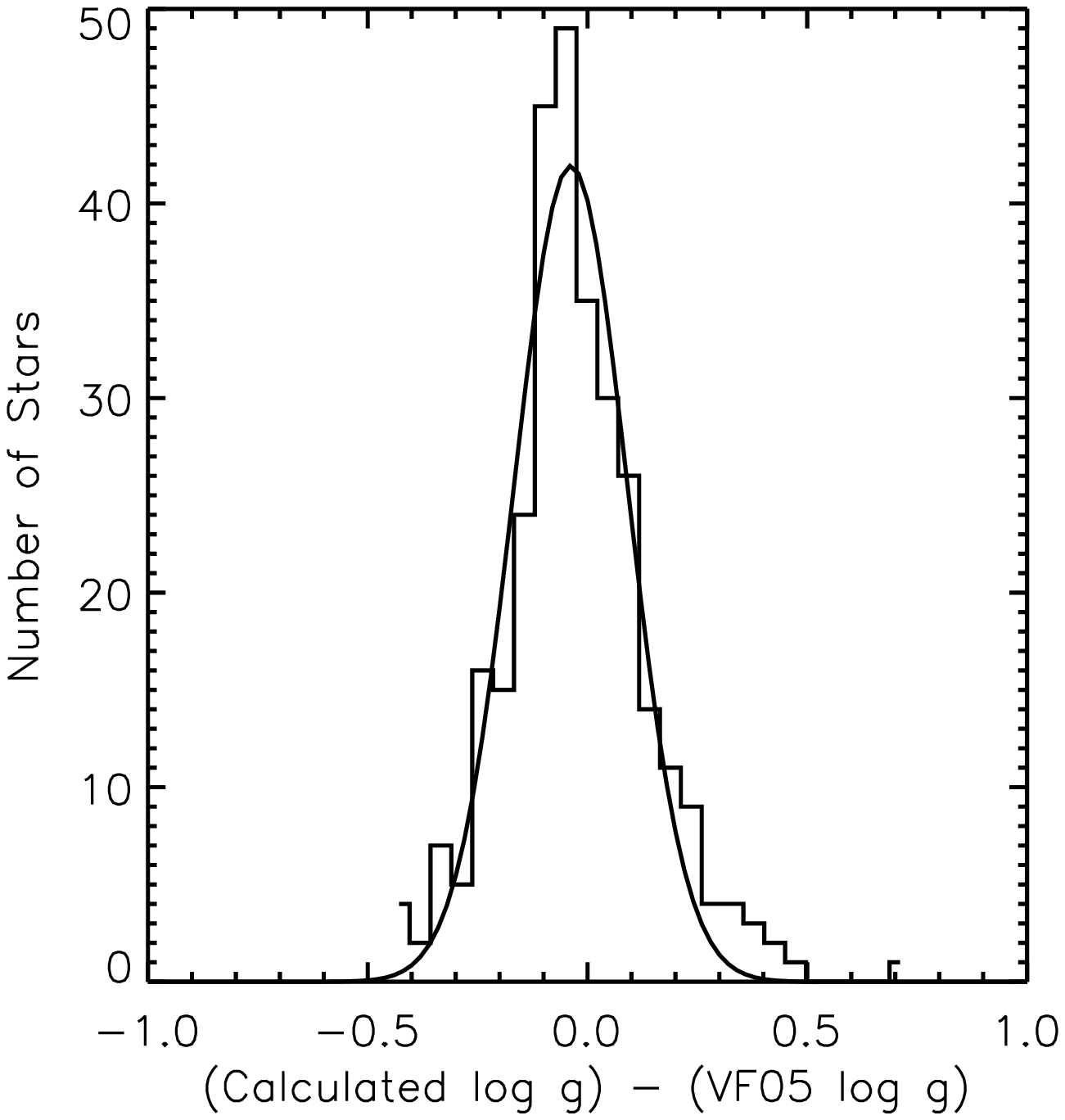}
\caption{Left: Scatter plot showing performance of $\log \, g$
calibration.  The solid line shows a theoretical perfect fit, and the
dotted lines show the calibration's 1-$\sigma$ error.  Although stars
with $\log \, g \geq 3.5$ formed the training set, this calibration
is not effective at separating subgiants and dwarfs; thus we report its
useful range as $4.0 < \log \, g < 5.1$.  Right: Histogram of the
residuals of the $\log \, g$ calibration.  The residuals are
well modeled by a Gaussian distribution with $\sigma = 0.13$ dex,
centered at -0.038 dex.}
\label{loggfig}
\end{figure}

$\log \, g$ is difficult to measure from low-resolution spectra
because on its own, it induces very little change in the appearance
of the star's broadest absorption features.  In main-sequence stars,
on which we focus our analysis, $\log \, g$ is tied to $T_{\rm eff}$
in a one-parameter sequence; however, Lick indices plus $T_{\rm
eff}$ do not give an adequate fit.  The Balmer lines, which are very
temperature-sensitive, nevertheless do broaden noticeably as $\log
\, g$ increases.  We therefore attempted to extract the $\log \,
g$ information from the Balmer lines by putting one nonlinear term
into the fit: $T_{\rm eff} \rm \left ( H\beta + H\gamma_F \right )$.
This produced an acceptable calibration with $\sigma_{\log \, g} =
0.13$\ dex and a reduced $\chi^2$ of 4.10.  The fit coefficients and
statistics are given in Table \ref{calibrations} and the calibration
and error histogram are plotted in Figure \ref{loggfig}.

\begin{figure}
\plotone{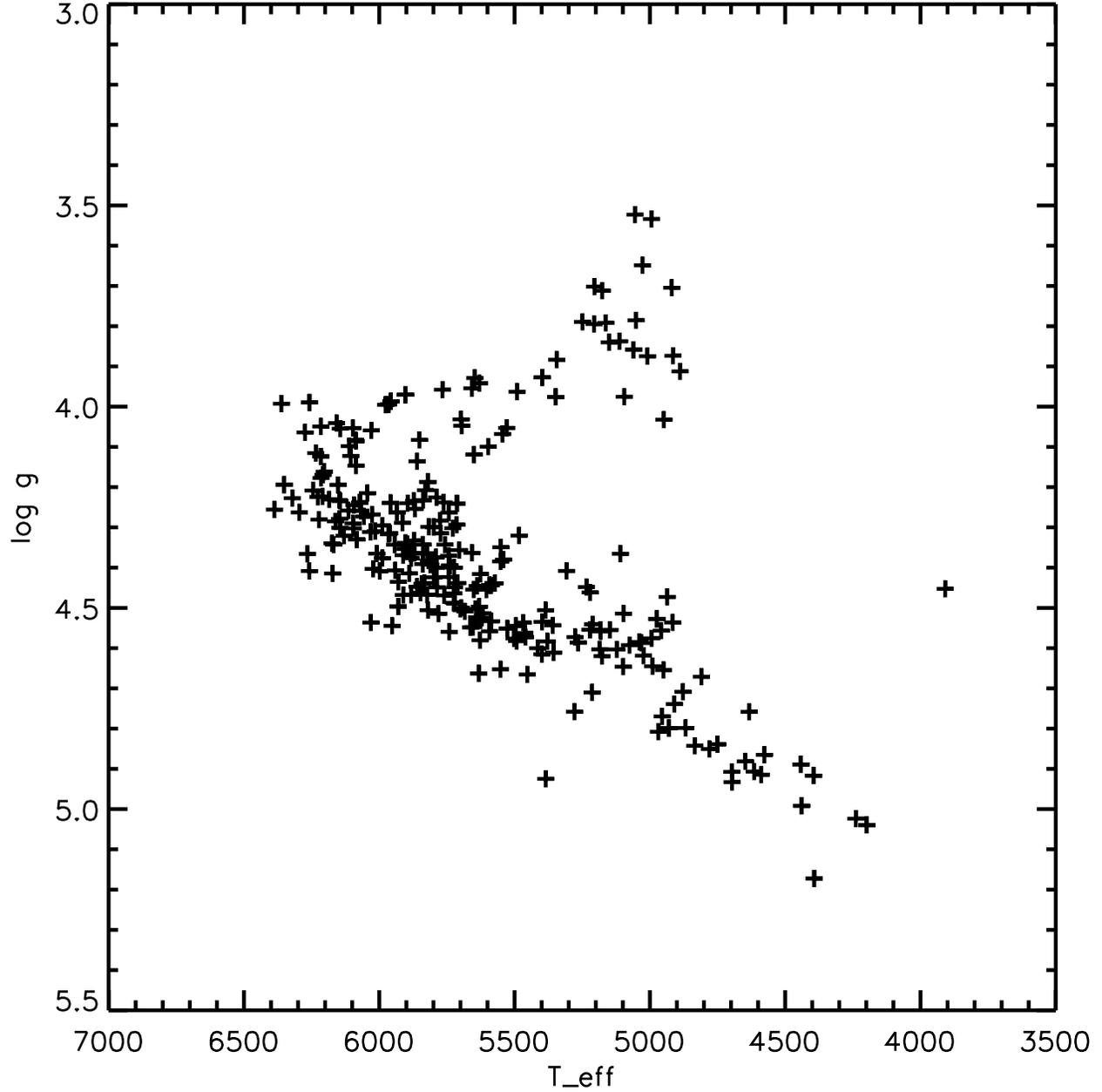}
\caption{Modified HR diagram of the VF05 stars in our dataset.
Instead of luminosity, we have plotted $\log \, g$ on the vertical
axis, showing that the divide between dwarfs and subgiants occurs at
$\log \, g \sim 4.0$\ dex.}
\label{modifiedhr}
\end{figure}

Figure \ref{modifiedhr}, a modified HR diagram of our training set with
$\log \, g$ on the vertical axis instead of luminosity, shows that
almost all FGK stars with $\log \, g < 4.0$ dex are on the subgiant
branch.  It is therefore clear from the scatter plot in Figure
\ref{loggfig} that our calibration does not distinguish subgiants from
dwarfs: all stars with $\log \, g < 4.0$ dex have their $\log \, g$
values scattered upward.  Either a larger training set with more
subgiants and giants or a separate calibration would be required to
cross the main-sequence turnoff.  We are still characterizing most of
the range in $\log \, g$ where planets have been found---the
planet-bearing star with the lowest $\log \, g$ to date, 3.78, is HD
27442, discovered by \cite{HD27442}.  However, if this method is
extended to fainter stars, care should be taken not to confuse distant
giants with nearer subgiants and dwarfs.

\subsection{Comparison with Previous Studies}
\label{litcompare}

We compared our calibrations with the result of inverting the W94
fitting functions for the stars in our training set, and with the
most recent calibrations between Str\"{o}mgren indices and stellar
atmospheric parameters, in \cite{sarah}.  \cite{BS94} also give fits
Fe5270$( \Theta, {\rm [Fe/H]}, \log \, g ) $ and H$\beta ( \Theta,
{\rm [Fe/H]}, \log \, g )$.  However, unless one of the independent
variables is fixed (for example, $T_{\rm eff}$ determined from
photometry), these fits cannot be inverted to solve for atmospheric
parameters, because the system would be underdetermined.

The following procedure was used to replicate the methods of WJ03 and
invert the W94 fitting functions:
\begin{enumerate}
\item We created a grid in $( T_{\rm eff},\ {\rm [Fe/H]},\ \log \,
g )$ space that encompasses the range $3570 < T_{\rm eff} < 6720$K
($1.4 > \Theta > 0.74$), $-1.5 < {\rm [Fe/H]} < 1.0\ {\rm dex}$ and
$0.0 < \log \, g < 8.0\ {\rm dex}$; with spacing $\Delta T_{\rm eff}
= 40$K, $\Delta {\rm [Fe/H] = 0.025\ dex}$, and $\Delta \log \, g =
0.05\ {\rm dex}$.  The grid spacing was selected to approximate the
precision of the VF05 data.
\item As in WJ03, we defined a figure of merit for one star as \[
G_x^2 = \sum_m (I_m - C_{m,x})^2 / \sigma^2_m , \] where $I_m$ is
the observed index, $\sigma_m$ is the index error (as given in Table
\ref{transformations}), and $C_{m,x}$ is the calculated index value
for the set of atmospheric parameters $x$.
\item $G_x$ was calculated for every point in the $( T_{\rm eff},\ {\rm
[Fe/H]},\ \log \, g )$ grid.  W94 give separate fitting functions for
cool and warm stars, which overlap where $5040 < T_{\rm eff} < 5160$K.
For gridpoints in this region, we use both sets of fitting functions
to calculate $G_x$ and retain the minimum of the two resulting values.
\item The minimum value of $G$ and its corresponding atmospheric
parameters were found.  This process was repeated for every star in
the data.
\end{enumerate}
A full grid search is computationally expensive, but it ensures
that the global minimum in $G$ is found for every star.  This is the
most numerically robust way to find a best-fit set of parameters.

\begin{figure}
\plotone{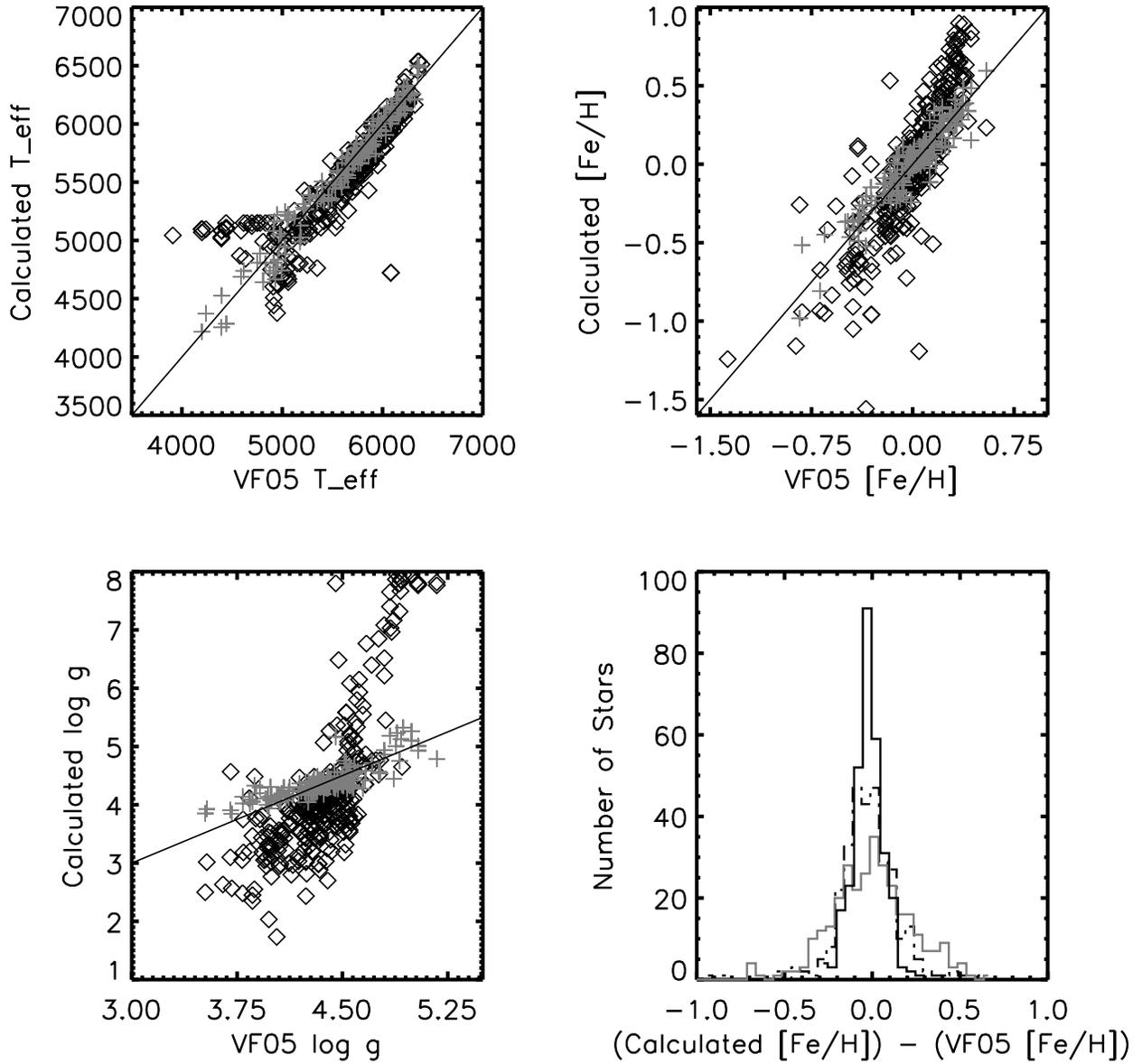}
\caption{Results of inverting W94 fitting functions.  Top left: $T_{\rm
eff}$; top right: [Fe/H], bottom left: $\log \, g$.  In each scatter
plot, the black diamonds plot atmospheric-parameter values calculated
using the WJ03 method against those of VF05.  For comparison, the
gray plus signs show the performance of the calibrations presented
in this paper.  The black, solid lines represent a 1:1 correlation.
In the lower right, we show histograms of the residuals between
metallicities determined by the WJ03 method and the VF05 values (solid
gray); the WJ03-method residuals if the slope of the calculated vs.
actual [Fe/H] trend is corrected to match the metallicity scale of VF05
(dotted black); and the residuals from our own calibration.}
\label{W94ff}
\end{figure}

In Figure \ref{W94ff}, we compare the results of the WJ03 method of
finding atmospheric parameters with the new calibrations presented
here.  For stars warmer than 5015K, the two methods do a comparable job
of finding effective temperature, though our calibrations show less
scatter.  For [Fe/H] and $\log \, g$, strong systematic differences
are evident between the the two methods.  The W94 metallicity scale
appears to have a steeper tilt than that of VF05: Stars more metal-rich
than the Sun have their metallicity substantially overpredicted by up
to 0.5 dex, while [Fe/H] is underpredicted in stars more metal-poor
than the Sun.  In addition, as the histogram in Figure \ref{W94ff}
shows, the [Fe/H] values found from the fitting functions are scattered
more widely than those determined by our own calibrations.  Since the
systematic WJ03-method trend in [Fe/H] is mainly linear, the slope
can be corrected to force the resulting [Fe/H] values to conform to
the metallicity scale of VF05; doing so reduces $\sigma_{[Fe/H]}$
from 0.21 dex to 0.12 dex.  The systematic trend in $\log \, g$
is even more pronounced, and has the added complication of being
nonlinear: Fitting the WJ03-method results for stars with $\log \,
g < 4.5$ dex would yield a different slope than for stars with $\log \,
g > 4.5$ dex.  Furthermore, there is substantial scatter in $\log \, g$
values found by inverting the W94 fitting functions for stars with
$\log \, g < 4.5$ dex.

Our $T_{\rm eff}$ calibration, $\sigma = 82$K, performs comparably to
that of \cite{sarah}, $\sigma = 80$K.  However, the benefits of using
low-resolution spectroscopy are most evident when measuring [Fe/H] and
$\log \, g$.  Both of these parameters have only subtle effects on the
shape of the stellar continuum, which can be measured well with
photometry.  Instead, [Fe/H] and $\log \, g$ change the depth and
profile of broad spectral features in ways that are reflected in Lick
indices.  True empirical estimators for $\log \, g$ are uncommon in the
literature; most methods are based on the combination of photometry and
stellar models (see, e.g., Lastennet et al. 2001).  One semiempirical
calibration, based on photometry in the Vilnius system, is the
\cite{vilnius} GK-giants calibration ($0.5 < \log \, g < 3.0$\ dex),
which has precision $\sigma = 0.3$ dex.  Our $\log \, g$ calibration
fills an important niche: It functions as a proof of concept for fully
empirical modeling of $\log \, g$ (which is possible now that
measurements from high-resolution spectra have become more precise and
uniform), and it demonstrates the appreciable sensitivity of
low-resolution spectra to $\log \, g$.  Our [Fe/H] calibration, with
$\sigma = 0.07$ dex, is more precise than that of \cite{sarah}
($\sigma_{\rm [Fe/H]} = 0.10$ dex) or \cite{SN89} ($\sigma_{\rm [Fe/H]}
= 0.13$ dex), and more accurate: Our Gaussian fit to the calibration
residuals is centered at -0.017 dex, as opposed to -0.027 dex for
\cite{sarah} or -0.049 dex for \cite{SN89}.

\section{$M_V$ and Distance Measurements}

If one were to survey stars without Hipparcos parallaxes, as
planet-search projects will certainly have to do within the next two
years, it would be convenient to be able to estimate distances from
low-resolution spectra.  Such distance estimates could also be useful in
screening photometric transit candidates emerging from projects such as
the OGLE survey (Szyma\'{n}ski 2005).  The distance modulus can be
calculated by interpolating stellar models to find the value of $M_V$
that matches a star's effective temperature, surface gravity and
metallicity. We used the Y$^2$ isochrones of \cite{Ysquared}, for Solar
abundance ratios and ages between 1 and 13 Gyr.  The isochrones for the
two metallicities surrounding the star's [Fe/H] were interpolated in the
$T_{\rm eff}$ and $\log \, g$ plane with a Gaussian-weighted sum of
$M_V$ for the seven points nearest the star in temperature and gravity.
We then interpolated the two resulting values of $M_V$ linearly in the
metallicity dimension the to find the star's absolute magnitude and
distance modulus.

\begin{figure}
\plottwo{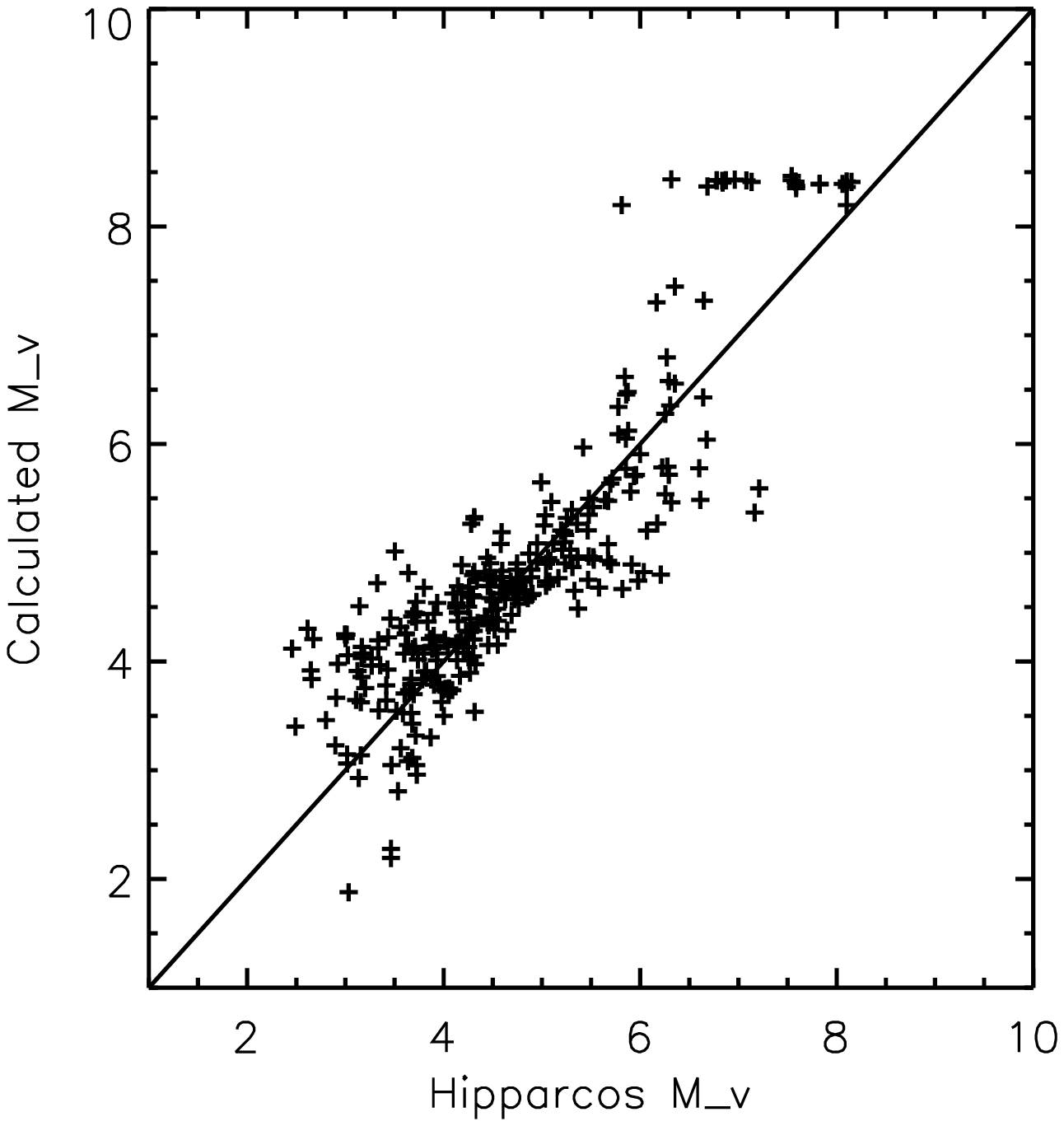}{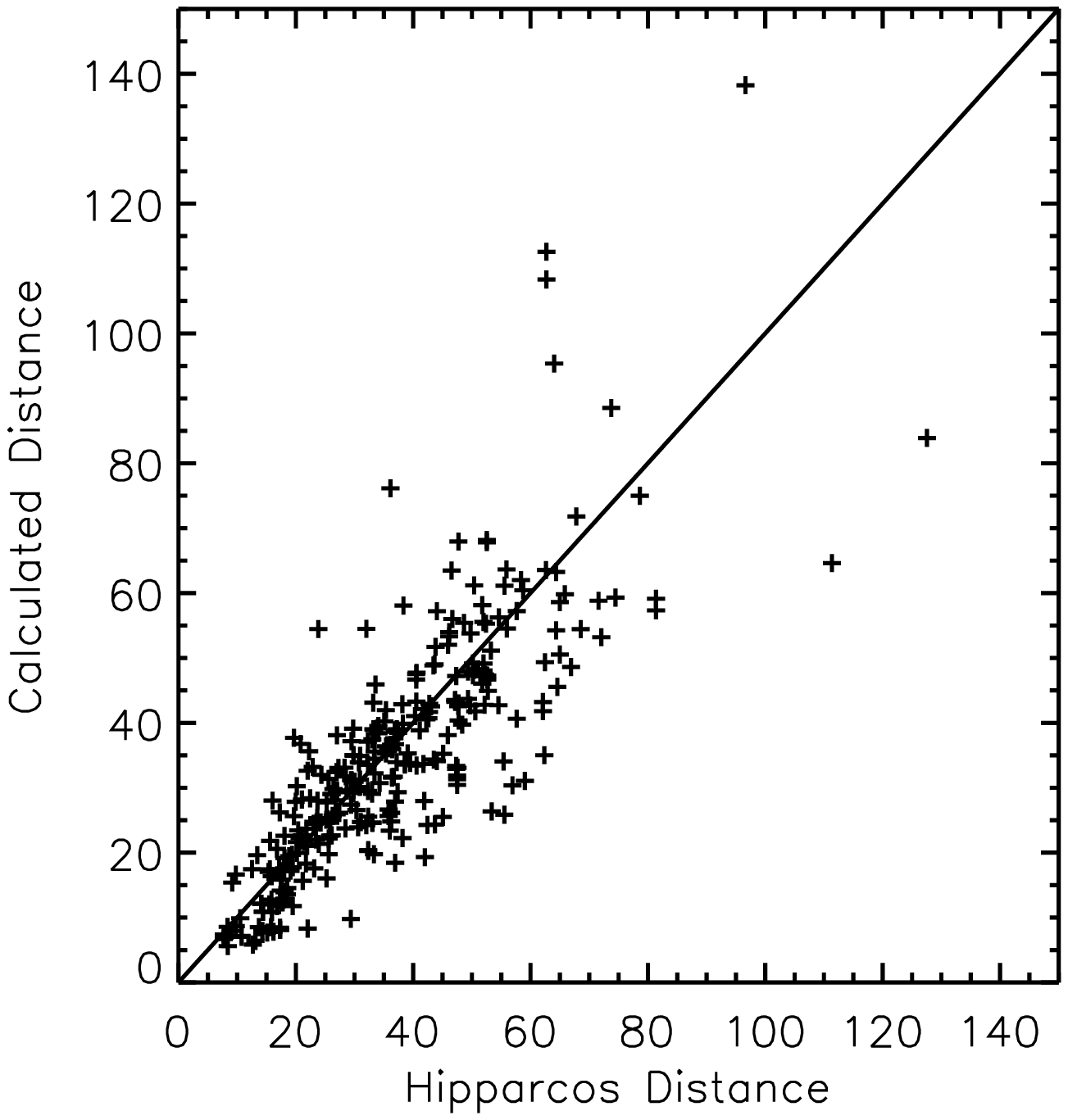}
\caption{Left: Comparison of M$_V$ estimates from our calibrations and
Y$^2$ isochrones to Hipparcos values.  Right: Comparison of estimated
distances and Hipparcos values.  In both plots, the solid line
represents a 1:1 correlation.}
\label{mvdist}
\end{figure}

Figure \ref{mvdist} compares the absolute magnitude and distances
estimated by our calibrations plus the Y$^2$ models with the Hipparcos
values.  This method gives distance measurements with uncertainty
$0.2d$, where $d$ is the actual distance to the star---we are
therefore estimating distance with $\sim 20\%$ accuracy.  As expected,
the limiting factor in our ability to spectroscopically estimate
distance is the accuracy of the $\log \, g$ calibration: the $M_V$
scatter plot reflects the increased uncertainty in $\log \, g$ for
subgiants and MSTO stars.  The $M_V$ estimate also has a ceiling at
6.5 magnitudes beyond which it is no longer linear.  This may reflect
the sparseness of isochrone data points in $T_{\rm eff}$ and $\log \,
g$ space for low-mass stars.

\section{Planeticity Calibration}

\cite{sarah} suggested the possibility of finding an empirical
calibration for ``planeticity,'' the presence or absence of a
detectable planet around a star.  Since the only known proxy for the
presence of a planet is the metallicity, the best chance of creating a
planeticity calibration should lie in finding trends in the abundance
ratios of planet-bearing stars.  One could then correlate Lick indices
with other abundance ratios besides [Fe/H] and use these ratios to
determine likely planeticity for each observed star.  Even a minimally
successful planeticity calibration would tremendously improve the
observing efficiency of Doppler surveys.

\begin{figure}
\plotone{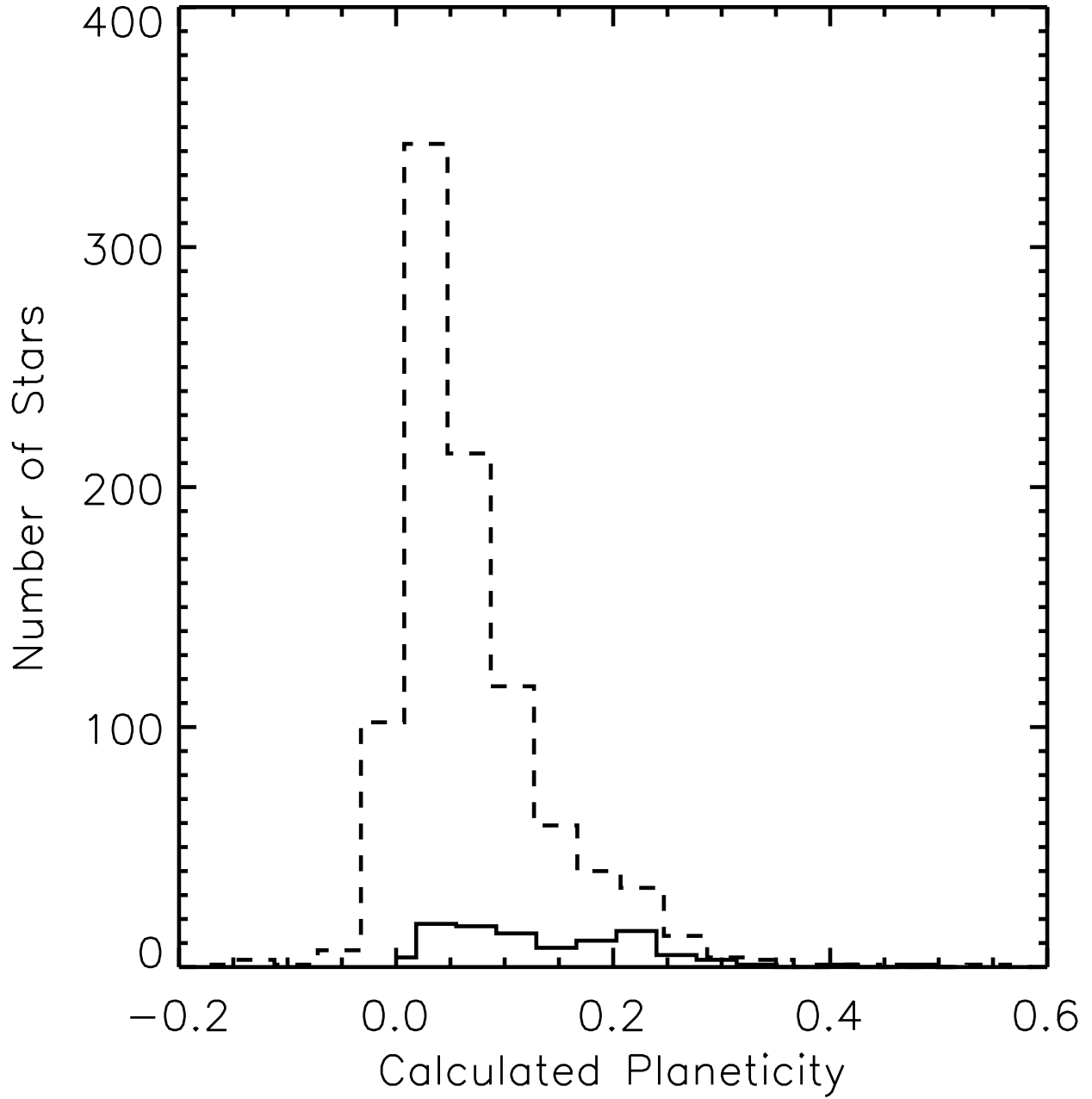}
\caption{Histograms of calculated planeticity for non-planet-bearing
stars (dashed line) and planet-bearing stars (solid line).  Stars
without planets were assigned a planeticity of 0, and stars with planets
were assigned a planeticity of 1.
The calibration is not able to distinguish stars with planets based
on metal abundances.}
\label{planeticity}
\end{figure}

\cite{sme} measured [Na/H], [Si/H], [Ti/H], [Fe/H] and [Ni/H],
referenced to $\log \, N_{Fe} = 7.50$, with other Solar abundances
from \cite{sun}.  Planets have been found around 98 of 1040 stars.
To form a calibration, we calculated each star's metal-to-Hydrogen
ratios as \[ M/H = {N_M \over N_H}\, , \] and each metal-to-metal
ratio as \[ M_1/M_2 = {M_1/H \over M_2/H}\, . \]  We then used
the Levenberg-Marquardt algorithm to find linear fits between the
abundance ratios and several properties of the known extrasolar
planets: $M \sin i$, $P$ and $M \sin i / P$ (a proxy for detectability,
since high-mass and short-period planets are the easiest to detect).
Stars without planets were assigned a planet mass of zero and a period
of 5000 days.  None of these fits were successful at replicating the
properties of the training set.  We also attempted a calibration aimed
at discerning whether a particular star would have any planet at all,
without reference to its mass or period: planet-bearing stars were
assigned a planeticity of one, and stars without planets were assigned
a planeticity of zero.  As shown in Figure \ref{planeticity}, this was
unsuccessful: although stars with planets had slightly higher mean
planeticity than those without, the calibration could not separate
the populations of planet- and non-planet hosts.

\begin{figure}
\plotone{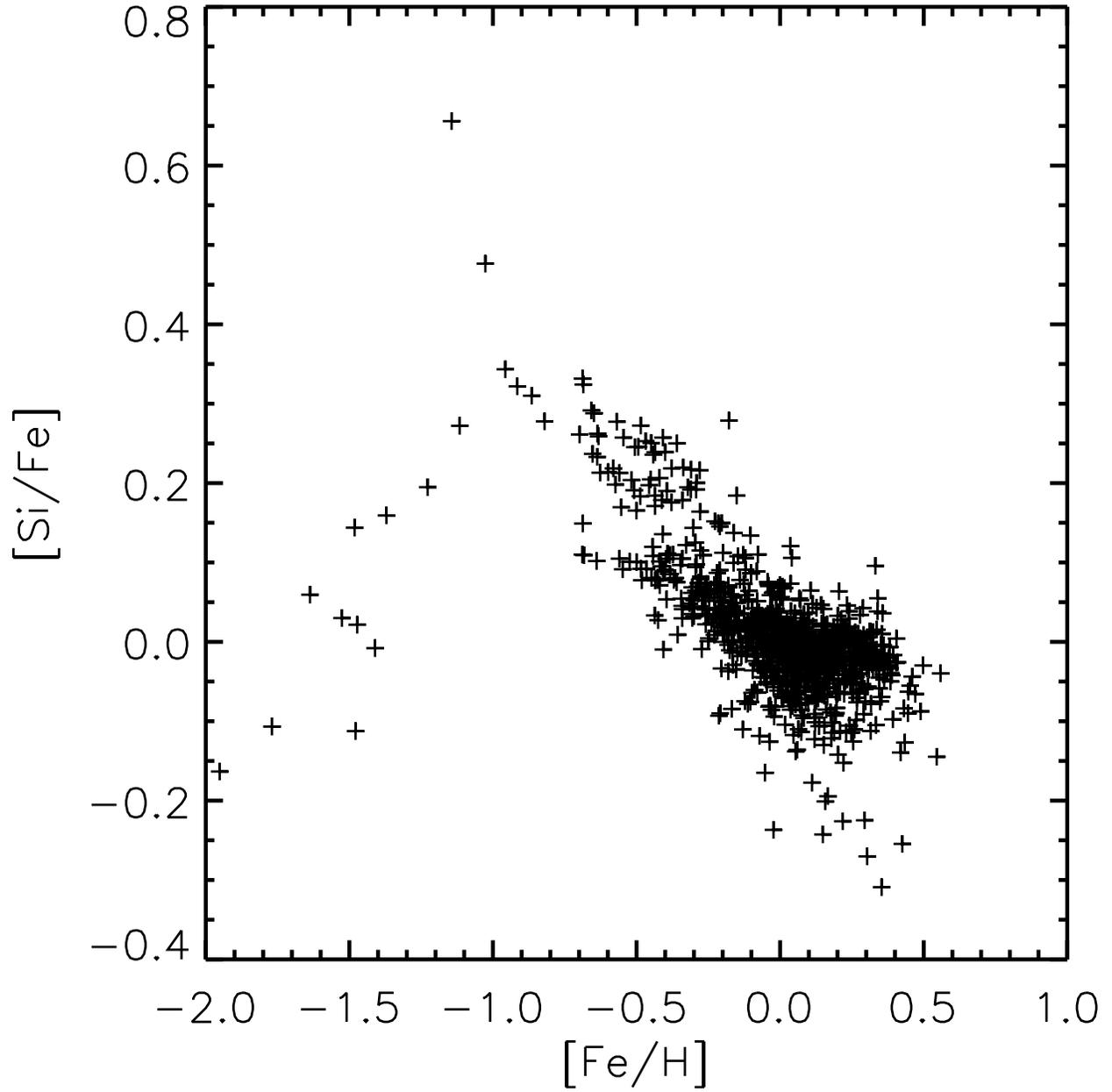}
\caption{[Si/Fe] vs. [Fe/H] for the stars in the VF05 data set.
None of the metal-rich stars show significant alpha enhancement.}
\label{SiFe}
\end{figure}

One implication of the core accretion model of planet formation
is that Oxygen-enhanced protostellar disks would have an enhanced
probability of producing giant planets, which are thought to have
cores composed mainly of H$_2$O ice.  Therefore, if a chemical predictor of planeticity
exists within the population of super-metal-rich stars, it is likely
[$\alpha$/Fe].  With abundance data that included two Fe-peak
elements (Fe and Ni) and two alpha elements (Si and Ti), we were
unable to distinguish planet hosts from other stars.  This might
imply that alpha enhancement is not correlated with planet formation.
It is also possible that a relationship between [$\alpha$/Fe] and
planet formation exists, but we could not find it because our stars
are of approximately Solar composition.  Figure \ref{SiFe} shows
[Si/Fe] vs. [Fe/H] for the \cite{sme} stars.  [Si/Fe] in stars above
Solar metallicity is confined to a tight locus that covers barely
a factor of two.  A more significant increase in alpha abundance
may be necessary to increase the probability of planet formation.
Finally, until terrestrial planets and planets at tens of AU from
their host stars can be detected, we must assume that many stars
that are presumed planetless do in fact have a system of satellites.
This could preclude the possibility of a planeticity calibration,
since these stars should not differ chemically from the stars that
are known planet hosts.  The dynamical interaction of the planets and
the disk would likely determine the final configuration of the system.



\section{Discussion}

Our Lick-indices method of measuring atmospheric parameters has the
advantage of being both precise and extremely efficient, requiring only
one observation per star.  Although small, systematic error trends no
doubt remain in our calibrations, these systematics should match those
contained in the VF05 dataset; our internal, random errors are small.
The [Fe/H] and $T_{\rm eff}$ calibrations are especially robust and
can be modestly extrapolated beyond the temperature and metallicity
ranges of the training set.  In producing our calibrations, we have
been struck by the utility of linear or low-order fits involving
several independent terms.  These fits are more trustworthy than
high-order fits on a few indices, which can be numerically unstable
to interpolations even within the parameter space covered by the
training-set data.  We speculate that inverting the W94 fitting
functions results in noticeable scatter of calculated atmospheric
parameters because the fitting functions are third-order polynomials
that may not be well behaved within the range of Lick indices analyzed
here.  Additionally, trial and error reveals that constructing a
numerically robust empirical fit requires at least 10 times as many
training-set members as terms in the calibration.  Adding one or two
thoughtfully chosen nonlinear terms, which can reasonably be expected to
carry information about the parameter being measured, is therefore a
far more economical way to make a fit more precise than increasing
the order of the fit, which would necessitate the addition of many
new training-set members (in this case, observations).

With our $T_{\rm eff}$ calibration, we are able to screen out candidate
stars hotter than the $\sim{\rm F}7$ temperature limit for combined
high-accuracy spectral synthesis modeling and $3-5 \: {\rm m s^{-1}}$
Doppler precision (Fischer \& Valenti 2005).  Our $\log \, g$
calibration is currently valid for FGK main-sequence dwarfs because
these stars comprise the bulk of the VF05 training set used to build the
calibrations.

Certainly, as Doppler velocity surveys are extended to include stars
that lack accurate Hipparcos parallaxes, it will be vital to extend
the $\log \, g$ calibration to subgiants and giants.  This extension
should be straightforward given an adequate training set, and would be
of considerable value.  It could also be used, for example, to select
stars in the Hertzsprung gap for inclusion in radial-velocity surveys
(see Sato et al. 2005b, Johnson 2005).  Stars that are evolving across
the Hertzsprung gap are intrinsically luminous and relatively rare, and
hence tend to lie outside the regime of good trigonometric parallaxes.
Characterization of such stars is, however, a matter of great current
interest for radial-velocity programs because many members of this
population were originally early F through late B stars when they were
on the main sequence.  By monitoring Hertzsprung gap stars with the
Doppler technique, one can probe the planetary distribution endemic
to stars in the range $1.5 M_{\odot}$ to $5 M_{\odot}$.  In any event,
our current $T_{\rm eff}$ and $\log \, g$ calibrations can be combined
with the $Y^2$ stellar models to give $20\%$-accurate distance estimates
for nearby stars.

A planeticity calibration tied to specific abundance ratios remains
a tantalizing idea, but we find that [Fe/H] remains the best
predictor of presence of a detectable planet.  By screening stars
for high [Fe/H] with our Lick indices calibration, we can use the
planet-metallicity correlation to increase the efficiency of Doppler
planet searches.  The N2K Consortium is obtaining low-resolution
spectra of 2000 high-metallicity candidate stars (Robinson et al. 2005,
in preparation), from which we expect to identify $\sim 500$ stars
with ${\rm [Fe/H]} \geq 0.2 \, {\rm dex}$.  This pool should yield
$\sim 30$ hot Jupiters and 2-3 transits of stars bright enough for
high-precision follow-up studies from both ground and space.

\acknowledgements
It is a pleasure to thank Peter Bodenheimer for advice on stellar
models and Greg Spear for observing assistance.  This work is based
on observations conducted at Lick Observatory and Kitt Peak National
Observatory.  S.R., J.S. and S.M.A. were supported by Fellowships
from the US National Science Foundation Graduate Research Fellowship
Program.  Additional support for this research was provided by the
NASA Origins of Solar Systems Program through grant NNG04GN30G to G.L.

Facilities: \facility{Lick Observatory(Nickel Spectrograph)},
\facility{KPNO 2.1m(GoldCam Spectrograph)}





\clearpage
\begin{deluxetable}{lrrrc}
\tabletypesize{\small}
\tablecaption{Matching the Lick system: Linear transformations from
observed to published Lick indices and index errors 
\label{transformations}}
\tablehead{
\colhead{Index} & \colhead{Slope} & \colhead{Intercept} &
\colhead{Error} & \colhead{N Rejected\tablenotemark{a}} }
\startdata
Ca4227\tablenotemark{b} & 1.170 & -0.081 & 0.216 & 2 \\
& 1.101 & -0.309 & 0.210 & 1 \\
G4300 & 1.182 & -0.854 & 0.439 & 2 \\
& 1.229 & -0.911 & 0.372 & 3 \\
HgF & 1.079 & -0.250 & 0.387 & 2 \\
& 1.068 & -0.019 & 0.520 & 3 \\
Fe4383 & 1.016 & 0.287 & 0.569 & 1 \\
& 0.975 & -0.575 & 0.693 & 1 \\
Fe4531 & 1.011 & -0.182 & 0.316 & 1 \\
& 0.988 & -0.401 & 0.402 & 1 \\
Fe4668 & 1.092 & -0.526 & 0.588 & 2 \\
& 1.067 & -0.234 & 0.607 & 2 \\
Hbeta & 1.021 & -0.200 & 0.236 & 2 \\
& 0.991 & -0.133 & 0.161 & 2 \\
Fe5015 & 1.198 & 0.066 & 0.580 & 0 \\
& 1.055 & -0.278 & 0.501 & 0 \\
Mg2 & 1.043 & 0.044 & 0.009 & 1 \\
& 1.036 & 0.033 & 0.010 & 1 \\
Mgb5177 & 1.490 & 0.590 & 0.298 & 1 \\
& 1.376 & 0.518 & 0.355 & 2 \\
Fe5270 & 1.203 & -0.021 & 0.234 & 1 \\
& 1.186 & -0.257 & 0.308 & 0 \\
Fe5406 & 1.348 & -0.215 & 0.247 & 1 \\
& 1.183 & -0.313 & 0.223 & 2 \\
Na5895 & 1.004 & 0.219 & 0.303 & 1 \\
& 1.149 & -0.279 & 0.211 & 2 \\
\enddata
\tablenotetext{a}{Number of points rejected from final computation of
transformation}
\tablenotetext{b}{Top row gives transformations for data taken at
the Nickel 1m telescope; bottom row gives transformations for data
taken at KPNO 2.1m telescope}
\end{deluxetable}

\clearpage
\begin{deluxetable}{rcccccccccccccc}
\tabletypesize{\scriptsize}
\tablewidth{0pt}
\rotate
\tablecaption{Lick Indices for Training-Set Stars.
\label{starindices}}
\tablehead{
\colhead{HD} & \colhead{Site\tablenotemark{a}} &
\colhead{Ca4227} & \colhead{G4300} & \colhead{H$\gamma_F$}
& \colhead{Fe4383} & \colhead{Fe4531} & \colhead{Fe4668}
& \colhead{H$\beta$} & \colhead{Fe5015} & \colhead{Mg$_2$} &
\colhead{Mg $b$} & \colhead{Fe5270} & \colhead{Fe5406} & \colhead{Na D}
}
\startdata
400\tablenotemark{b} & N &  0.482 & 1.13 & 2.13 &  0.975 & 1.88 & 1.24 & 3.42 & 3.37 & 0.0696 & 1.29 & 1.34 &  0.352 & 1.04 \\
  & N &  0.482 & 1.13 & 2.13 &  0.975 & 1.88 & 1.24 & 3.42 & 3.37 & 0.0696 & 1.29 & 1.34 &  0.352 & 1.04 \\
691 & N &  0.977 & 5.51 &  -1.69 & 4.63 & 3.01 & 4.98 & 2.19 & 6.09 &  0.198 & 3.95 & 3.43 & 2.09 & 2.26 \\
  & N &  0.977 & 5.51 &  -1.69 & 4.63 & 3.01 & 4.98 & 2.19 & 6.09 &  0.198 & 3.95 & 3.43 & 2.09 & 2.26 \\
3079 & N &  0.477 & 2.84 & 1.54 & 1.61 & 1.99 & 1.77 & 3.05 & 4.15 & 0.0828 & 1.66 & 1.68 &  0.716 & 1.08 \\
  & N &  0.477 & 2.84 & 1.54 & 1.61 & 1.99 & 1.77 & 3.05 & 4.15 & 0.0828 & 1.66 & 1.68 &  0.716 & 1.08 \\
  & K &  0.470 & 3.59 & 1.05 & 2.79 & 2.30 & 1.58 & 3.15 & 3.46 & 0.0807 & 1.52 & 1.35 &  0.516 &  0.861 \\
3765 & K & 2.99 & 5.83 &  -1.58 & 8.63 & 4.36 & 6.27 & 1.06 & 6.33 &  0.386 & 6.86 & 4.19 & 2.82 & 4.70 \\
3770 & K &  0.410 & 4.22 & 1.18 & 2.58 & 2.36 & 1.79 & 3.16 & 3.68 & 0.0759 &  0.905 & 1.26 &  0.493 & 1.16 \\
4256 & K & 3.81 & 5.71 &  -1.77 & 8.23 & 4.84 & 7.36 & 1.80 & 6.49 &  0.458 & 8.02 & 4.33 & 3.25 & 6.20 \\
  & K & 3.73 & 5.96 &  -1.91 & 9.61 & 4.86 & 7.17 &  0.857 & 6.53 &  0.459 & 8.11 & 4.36 & 3.26 & 6.23 \\
4903 & K &  0.305 & 3.75 & 1.25 & 2.50 & 2.44 & 1.84 & 3.30 & 3.52 & 0.0763 & 1.27 & 1.30 &  0.534 &  0.959 \\
5470 & K &  0.625 & 4.81 &  0.268 & 3.85 & 2.88 & 4.26 & 3.13 & 4.81 &  0.117 & 2.13 & 1.99 &  0.848 & 1.62 \\
6963 & K & 1.28 & 5.03 &  -1.83 & 4.91 & 2.84 & 2.73 & 2.00 & 4.13 &  0.157 & 3.40 & 2.31 & 1.33 & 1.75 \\
7590 & K &  0.694 & 3.87 &  0.531 & 2.90 & 2.43 & 1.90 & 2.82 & 3.71 & 0.0946 & 1.96 & 1.52 &  0.704 & 1.12 \\
8331 & K &  0.640 & 5.73 &  -1.12 & 4.09 & 2.76 & 3.43 & 2.44 & 4.34 &  0.117 & 1.96 & 1.90 &  0.923 & 1.31 \\
9070 & K & 1.11 & 5.20 &  -1.21 & 5.51 & 3.28 & 5.75 & 2.73 & 5.65 &  0.170 & 3.04 & 3.07 & 1.41 & 2.39 \\
9331 & K & 1.20 & 5.59 &  -1.50 & 5.58 & 3.31 & 5.20 & 2.56 & 5.36 &  0.169 & 3.37 & 2.69 & 1.35 & 1.98 \\
10086 & K & 1.07 & 4.66 &  -1.17 & 4.67 & 2.99 & 4.11 & 2.47 & 4.78 &  0.147 & 2.78 & 2.27 & 1.22 & 1.68 \\
11850 & K & 1.26 & 5.12 &  -1.47 & 4.85 & 3.01 & 3.74 & 2.19 & 4.67 &  0.152 & 2.97 & 2.43 & 1.25 & 1.73 \\
  & K & 1.20 & 5.26 &  -1.29 & 4.89 & 3.04 & 3.86 & 2.36 & 4.65 &  0.155 & 2.95 & 2.56 & 1.30 & 1.89 \\
12235 & K &  0.512 & 4.67 &  0.515 & 3.51 & 2.90 & 4.33 & 3.24 & 4.74 &  0.111 & 1.82 & 2.03 &  0.743 & 1.61 \\
12328 & K & 1.62 & 5.85 &  -1.69 & 6.87 & 3.72 & 6.11 & 1.30 & 5.60 &  0.248 & 4.82 & 3.55 & 2.17 & 2.70 \\
12414 & K &  0.227 & 2.42 & 2.72 &  0.902 & 1.69 &  0.809 & 3.63 & 2.24 & 0.0554 &  0.980 &  0.972 &  0.276 &  0.899 \\
12661 & K &  0.662 & 5.36 &  -1.33 & 5.29 & 3.22 & 6.80 & 2.73 & 5.67 &  0.180 & 3.34 & 2.64 & 1.34 & 2.31 \\
12846 & K &  0.824 & 5.15 &  -1.15 & 3.83 & 2.35 & 2.32 & 2.28 & 3.59 &  0.129 & 2.70 & 1.71 &  0.798 & 1.38 \\
13531 & K & 1.19 & 5.18 &  -1.45 & 4.50 & 3.03 & 3.38 & 2.09 & 4.53 &  0.150 & 2.83 & 2.31 & 1.25 & 1.77 \\
13825 & K &  0.708 & 5.37 &  -1.39 & 5.00 & 3.17 & 5.29 & 2.58 & 5.10 &  0.174 & 2.96 & 2.40 & 1.17 & 2.08 \\
16275 & K &  0.826 & 4.67 & -0.482 & 4.72 & 3.14 & 5.43 & 2.87 & 5.34 &  0.140 & 2.96 & 2.61 & 1.18 & 2.18 \\
17230 & K & 6.86 & 5.55 &  -1.58 & 8.34 & 5.40 &  -3.20 & 1.45 & 6.32 &  0.572 & 8.53 & 5.28 & 4.14 & 11.3 \\
  & K & 6.78 & 5.48 &  -1.44 & 8.09 & 5.57 &  -3.07 & 1.27 & 6.32 &  0.569 & 8.38 & 5.27 & 4.19 & 11.3 \\
18143 & K & 2.35 & 5.55 &  -1.69 & 8.17 & 4.11 & 7.39 & 1.45 & 6.40 &  0.344 & 5.76 & 3.74 & 2.48 & 4.41 \\
18803 & K &  0.911 & 4.90 &  -1.15 & 3.74 & 3.01 & 4.57 & 2.47 & 4.51 &  0.170 & 3.71 & 2.90 & 1.24 & 2.27 \\
19373 & K &  0.587 & 4.26 &  0.727 & 2.30 & 2.62 & 2.96 & 3.09 & 4.28 &  0.104 & 2.01 & 1.90 &  0.755 & 1.55 \\
21019 & K &  0.526 & 5.63 &  -1.71 & 3.61 & 2.36 & 2.03 & 1.73 & 3.48 &  0.105 & 1.83 & 1.54 &  0.765 & 1.38 \\
21197 & K & 5.97 & 5.45 &  -1.59 & 8.79 & 5.65 &  -2.00 & 1.68 & 6.91 &  0.556 & 9.12 & 5.58 & 4.15 & 10.5 \\
22072 & K & 1.06 & 6.07 &  -1.99 & 4.57 & 3.09 & 4.29 & 1.28 & 4.65 &  0.207 & 3.87 & 2.53 & 1.47 & 1.84 \\
23249 & K & 1.19 & 6.14 &  -1.66 & 6.96 & 3.69 & 7.76 & 1.51 & 5.77 &  0.257 & 3.98 & 3.50 & 2.14 & 3.25 \\
23596 & K &  0.542 & 4.63 &  0.527 & 3.56 & 2.93 & 3.97 & 3.34 & 4.73 &  0.110 & 1.62 & 1.83 &  0.794 & 1.53 \\
  & K &  0.529 & 4.61 &  0.476 & 3.96 & 2.96 & 3.72 & 3.31 & 4.74 &  0.109 & 1.51 & 2.03 &  0.746 & 1.57 \\
  & K &  0.575 & 4.66 &  0.529 & 3.53 & 2.66 & 3.73 & 3.26 & 4.70 &  0.106 & 1.83 & 2.01 &  0.863 & 1.74 \\
24365 & K &  0.809 & 6.37 &  -2.91 & 5.17 & 3.04 & 4.01 & 1.57 & 4.23 &  0.152 & 2.36 & 2.47 & 1.17 & 1.69 \\
24916 & K & 5.88 & 5.32 &  -1.26 & 8.41 & 5.39 &  -2.16 & 1.67 & 5.56 &  0.498 & 8.16 & 5.34 & 3.75 & 7.46 \\
25069 & K &  0.954 & 6.10 &  -1.87 & 6.53 & 3.84 & 7.22 & 1.48 & 5.89 &  0.239 & 4.00 & 3.68 & 2.29 & 3.00 \\
25790 & K &  0.611 & 5.47 &  -1.46 & 4.54 & 3.06 & 4.89 & 2.21 & 4.62 &  0.129 & 2.27 & 2.32 & 1.16 & 1.88 \\
26794 & K & 3.67 & 5.51 &  -1.61 & 8.15 & 4.17 & 5.56 &  0.705 & 5.85 &  0.443 & 8.84 & 4.29 & 2.87 & 5.42 \\
28005 & K &  0.477 & 5.12 & -0.714 & 4.47 & 3.16 & 5.84 & 2.87 & 5.54 &  0.156 & 2.71 & 2.28 & 1.14 & 2.17 \\
30508 & K &  0.838 & 6.38 &  -2.58 & 5.33 & 3.14 & 4.87 & 1.64 & 4.45 &  0.159 & 2.54 & 2.54 & 1.40 & 1.98 \\
30825 & K &  0.825 & 6.28 &  -1.62 & 4.71 & 3.14 & 3.85 & 1.44 & 4.64 &  0.158 & 2.53 & 2.39 & 1.40 & 1.72 \\
34575 & K &  0.955 & 5.32 &  -1.93 & 5.80 & 3.41 & 6.55 & 2.39 & 5.39 &  0.192 & 3.20 & 3.04 & 1.53 & 2.59 \\
52456 & N & 2.34 & 6.64 &  -1.78 & 7.05 & 3.25 & 5.07 & 1.41 & 5.74 &  0.282 & 5.35 & 3.56 & 2.38 & 3.41 \\
52711 & N & 1.03 & 4.44 &  0.374 & 2.99 & 2.54 & 2.20 & 2.69 & 3.83 &  0.110 & 2.37 & 1.75 &  0.745 & 1.37 \\
56124 & N & 1.05 & 4.58 &  -0.0229 & 3.65 & 2.58 & 2.58 & 2.75 & 3.89 &  0.121 & 2.41 & 1.96 &  0.991 & 1.50 \\
56303 & N &  0.982 & 4.36 &  0.697 & 3.43 & 2.49 & 3.44 & 3.07 & 4.34 &  0.111 & 2.15 & 2.10 &  0.840 & 1.44 \\
58781 & N & 1.24 & 5.67 &  -1.45 & 4.81 & 3.05 & 4.75 & 2.38 & 4.87 &  0.177 & 3.04 & 2.72 & 1.17 & 2.21 \\
59747 & N & 2.84 & 7.34 &  -2.06 & 7.03 & 3.79 & -0.526 & 1.29 & 3.92 &  0.299 & 5.93 & 4.01 & 2.60 & 3.24 \\
63433 & N & 1.39 & 5.13 & -0.598 & 4.44 & 2.73 & 3.24 & 2.27 & 4.58 &  0.143 & 2.78 & 2.47 & 1.28 & 1.69 \\
64468 & N & 3.44 & 6.84 &  -2.10 & 8.73 & 4.14 &  0.925 & 1.03 & 5.95 &  0.447 & 7.87 & 4.80 & 2.89 & 5.02 \\
65430 & N & 2.07 & 6.84 &  -2.17 & 5.92 & 3.24 & 4.96 & 1.57 & 5.21 &  0.303 & 6.46 & 3.19 & 1.73 & 3.09 \\
65583 & N & 1.70 & 5.67 &  -1.97 & 4.04 & 2.18 & 0.0508 & 1.58 & 3.46 &  0.206 & 4.96 & 2.03 &  0.797 & 1.91 \\
67767 & N &  0.909 & 6.12 &  -2.03 & 5.05 & 2.80 & 4.44 & 1.82 & 5.02 &  0.163 & 3.06 & 2.50 & 1.40 & 1.84 \\
68017 & N & 1.12 & 5.35 &  -1.14 & 3.71 & 2.04 & 2.21 & 2.07 & 3.30 &  0.154 & 3.43 & 1.71 &  0.599 & 1.52 \\
69809 & N & 1.06 & 5.24 & -0.265 & 4.25 & 2.79 & 5.65 & 2.96 & 5.27 &  0.150 & 2.87 & 2.52 & 1.04 & 1.98 \\
70843 & N &  0.612 & 3.35 & 2.33 & 2.20 & 2.20 & 2.58 & 3.80 & 4.21 & 0.0856 & 1.34 & 1.80 &  0.661 & 1.39 \\
72780 & N &  0.683 & 2.70 & 2.42 & 2.33 & 2.17 & 1.81 & 3.62 & 4.21 & 0.0844 & 1.32 & 1.86 &  0.621 & 1.26 \\
73344 & N &  0.723 & 3.66 & 1.58 & 2.52 & 2.36 & 2.57 & 3.35 & 3.87 & 0.0925 & 1.78 & 1.72 &  0.653 & 1.25 \\
73667 & N & 2.77 & 6.39 &  -1.83 & 5.73 & 2.98 & 1.99 &  0.981 & 4.45 &  0.343 & 7.34 & 2.91 & 1.90 & 3.04 \\
73668 & N &  0.929 & 4.11 &  0.684 & 3.04 & 2.50 & 2.44 & 3.02 & 3.95 &  0.107 & 2.01 & 1.98 &  0.931 & 1.67 \\
74156 & N &  0.729 & 4.00 & 1.07 & 2.53 & 2.31 & 3.01 & 3.30 & 4.48 & 0.0997 & 1.98 & 1.92 &  0.974 & 1.64 \\
75302 & N & 1.07 & 5.04 & -0.748 & 4.46 & 2.80 & 3.60 & 2.56 & 4.70 &  0.148 & 2.64 & 2.63 & 1.43 & 1.89 \\
75332 & N &  0.540 & 2.92 & 2.01 & 2.27 & 2.11 & 1.52 & 3.60 & 4.39 & 0.0895 & 1.75 & 1.80 &  0.678 & 1.35 \\
75732 & N & 1.06 & 6.57 &  -1.43 & 6.91 & 3.47 & 8.94 & 1.83 & 6.65 &  0.308 & 5.37 & 3.80 & 2.29 & 3.90 \\
75782 & N &  0.323 & 4.60 &  0.802 & 2.90 & 2.87 & 3.73 & 3.22 & 5.03 &  0.104 & 1.56 & 2.06 &  0.845 & 1.37 \\
76909 & N &  0.188 & 5.21 & -0.941 & 4.11 & 3.01 & 7.30 & 2.85 & 6.25 &  0.184 & 3.71 & 2.74 & 1.33 & 2.42 \\
80367 & N & 2.60 & 6.47 &  -1.68 & 7.20 & 3.70 & 5.66 & 1.20 & 5.51 &  0.349 & 7.77 & 3.86 & 2.56 & 3.39 \\
80606 & N & 1.03 & 5.79 &  -1.43 & 6.15 & 3.08 & 7.56 & 2.45 & 4.04 &  0.225 & 3.94 & 3.22 & 1.73 & 2.81 \\
82106 & N & 4.29 & 6.79 &  -1.67 & 9.06 & 4.70 & -0.726 & 1.50 & 3.60 &  0.428 & 8.25 & 4.70 & 3.32 & 4.81 \\
87836 & N &  0.794 & 5.80 & -0.709 & 4.00 & 3.07 & 6.72 & 2.77 & 5.67 &  0.164 & 2.73 & 2.69 & 1.42 & 1.87 \\
87883 & N & 3.11 & 6.55 &  -1.53 & 8.55 & 3.90 & 5.62 &  0.882 & 6.23 &  0.408 & 7.96 & 4.64 & 2.94 & 4.62 \\
88371 & N & 1.05 & 5.19 & -0.426 & 3.14 & 2.94 & 2.99 & 2.43 & 3.82 &  0.152 & 3.23 & 1.86 &  0.826 & 1.35 \\
88986 & N &  0.819 & 5.21 & -0.108 & 3.06 & 2.87 & 3.53 & 2.58 & 4.37 &  0.117 & 2.39 & 1.84 &  0.931 & 1.37 \\
89269 & N & 1.07 & 5.33 & -0.959 & 4.11 & 2.61 & 2.76 & 1.88 & 3.96 &  0.142 & 3.09 & 2.12 &  0.987 & 1.52 \\
89307 & N &  0.744 & 4.03 &  0.694 & 2.51 & 2.29 & 2.15 & 2.95 & 3.73 &  0.103 & 2.19 & 1.69 &  0.825 & 1.18 \\
91204 & N &  0.697 & 4.94 &  0.235 & 3.42 & 2.63 & 4.34 & 2.94 & 4.79 &  0.127 & 2.62 & 2.32 & 1.13 & 1.74 \\
95128 & N &  0.831 & 4.89 &  0.318 & 3.09 & 2.66 & 3.05 & 2.62 & 3.76 &  0.110 & 2.40 & 1.74 &  0.771 & 1.08 \\
96418 & N &  0.453 & 2.85 & 2.62 & 1.25 & 2.31 & 1.33 & 3.56 & 3.86 & 0.0729 & 1.11 & 1.51 &  0.548 &  0.911 \\
96574 & N &  0.774 & 3.66 & 1.78 & 1.95 & 2.27 & 2.15 & 3.41 & 3.78 & 0.0866 & 1.66 & 1.59 &  0.890 & 1.02 \\
97004 & N &  0.743 & 5.42 &  -1.22 & 5.52 & 3.61 & 7.57 & 2.46 & 5.84 &  0.214 & 4.39 & 3.18 & 1.47 & 2.45 \\
98388 & N &  0.449 & 2.06 & 2.87 & 1.77 & 2.17 & 1.49 & 3.80 & 3.89 & 0.0786 & 1.38 & 1.54 &  0.505 &  0.899 \\
98697 & N &  0.398 & 3.29 & 1.87 & 1.44 & 2.08 & 1.67 & 3.33 & 3.44 & 0.0741 & 1.09 & 1.51 &  0.434 & 1.06 \\
99491 & N & 1.25 & 5.94 &  -2.08 & 5.60 & 3.20 & 7.69 & 2.11 & 5.93 &  0.231 & 4.67 & 3.30 & 1.65 & 2.78 \\
99492 & N & 3.59 & 6.78 &  -1.87 & 9.51 & 4.76 & 7.58 &  0.730 & 6.63 &  0.461 & 8.87 & 4.81 & 3.59 & 5.95 \\
100180 & N &  0.576 & 4.01 &  0.865 & 2.65 & 2.42 & 2.20 & 2.89 & 3.92 & 0.0957 & 1.97 & 1.62 &  0.654 & 1.13 \\
102158 & N &  0.777 & 4.94 & -0.299 & 2.87 & 2.11 & 1.20 & 2.16 & 2.46 &  0.111 & 2.52 & 1.37 &  0.423 &  0.999 \\
103095 & N & 1.91 & 5.05 & -0.409 & 3.44 & 1.69 & -0.143 &  0.998 & 1.22 &  0.179 & 4.34 & 1.71 &  0.614 & 1.12 \\
103432 & N &  0.835 & 5.05 &  -1.10 & 4.09 & 3.08 & 3.05 & 2.24 & 4.52 &  0.145 & 3.38 & 2.74 & 1.30 & 1.64 \\
104556 & N & 1.24 & 6.78 &  -2.49 & 4.33 & 3.06 & 3.50 & 1.01 & 4.28 &  0.197 & 4.30 & 2.32 & 1.24 & 1.47 \\
104800 & N &  0.944 & 4.63 &  0.261 & 2.01 & 2.09 &  0.536 & 2.42 & 2.42 &  0.108 & 2.50 & 1.05 &  -0.0899 &  0.922 \\
105405 & N &  0.574 & 3.08 & 2.01 & 1.49 & 1.96 & 1.10 & 3.35 & 3.44 & 0.0761 & 1.47 & 1.27 &  0.580 & 1.03 \\
105631 & N & 1.43 & 5.88 &  -1.95 & 6.31 & 3.65 & 6.01 & 1.89 & 5.74 &  0.227 & 4.36 & 3.58 & 2.11 & 2.59 \\
106156 & N & 1.66 & 6.54 &  -2.26 & 6.46 & 3.54 & 6.53 & 1.91 & 5.43 &  0.238 & 4.71 & 3.56 & 1.86 & 2.66 \\
106252 & N &  0.822 & 4.81 & 0.0376 & 2.92 & 2.67 & 2.63 & 2.70 & 3.98 &  0.109 & 2.09 & 1.69 &  0.913 & 1.17 \\
106423 & N &  0.737 & 4.47 &  0.680 & 3.34 & 2.87 & 4.89 & 3.30 & 5.63 &  0.120 & 2.13 & 2.47 & 1.03 & 1.53 \\
107146 & N & 1.08 & 4.77 & 0.0711 & 3.36 & 2.93 & 2.96 & 2.52 & 4.14 &  0.118 & 2.19 & 1.86 & 1.11 & 1.23 \\
107213 & N &  0.476 & 3.29 & 2.03 & 1.41 & 2.60 & 2.98 & 3.63 & 4.31 & 0.0711 & 1.39 & 1.55 &  0.481 &  0.914 \\
107705 & N &  0.830 & 3.18 & 1.72 & 2.43 & 2.53 & 2.53 & 3.18 & 3.80 & 0.0805 & 1.44 & 1.65 &  0.590 &  0.979 \\
108874 & N &  0.898 & 5.74 &  -1.78 & 5.26 & 2.76 & 6.30 & 2.31 & 5.42 &  0.193 & 3.83 & 2.96 & 1.52 & 2.24 \\
109358 & N &  0.713 & 4.49 &  0.102 & 2.20 & 2.21 & 1.97 & 2.64 & 3.47 & 0.0971 & 1.95 & 1.43 &  0.665 & 1.06 \\
110315 & N & 5.66 & 6.75 &  -2.06 & 8.58 & 5.22 & 2.51 & 1.34 & 6.08 &  0.608 & 10.5 & 4.92 & 3.63 & 7.07 \\
111066 & N &  0.597 & 3.62 & 1.47 & 2.05 & 2.07 & 1.93 & 3.40 & 3.83 & 0.0797 & 1.55 & 1.54 &  0.603 &  0.955 \\
111395 & N & 1.47 & 5.32 &  -1.04 & 4.63 & 2.98 & 4.07 & 2.01 & 4.26 &  0.141 & 3.00 & 2.28 & 1.09 & 1.56 \\
111398 & N &  0.944 & 5.81 & -0.720 & 3.64 & 2.83 & 4.18 & 2.51 & 4.89 &  0.146 & 2.66 & 2.42 & 1.16 & 1.50 \\
111515 & N & 1.22 & 5.93 &  -1.41 & 3.38 & 2.58 & 1.23 & 1.60 & 3.33 &  0.177 & 4.10 & 2.04 &  0.746 & 1.59 \\
112060 & N &  0.814 & 6.44 &  -2.00 & 4.91 & 3.21 & 6.26 & 1.86 & 4.91 &  0.146 & 2.54 & 2.44 & 1.10 & 1.49 \\
114174 & N &  0.900 & 5.22 & -0.598 & 3.64 & 2.82 & 3.52 & 2.34 & 4.44 &  0.142 & 2.75 & 2.30 & 1.17 & 1.59 \\
114762 & N &  0.493 & 3.95 &  0.905 & 1.93 & 1.80 &  0.339 & 2.54 & 2.41 & 0.0811 & 1.77 & 1.07 &  0.239 &  0.988 \\
116442 & N & 1.64 & 6.32 &  -1.67 & 5.07 & 3.17 & 2.09 & 1.15 & 3.23 &  0.241 & 5.74 & 2.69 & 1.79 & 2.38 \\
116443 & N & 2.14 & 5.99 &  -1.79 & 6.38 & 3.26 & 2.53 &  0.885 & 4.38 &  0.314 & 6.88 & 3.23 & 1.90 & 2.78 \\
117126 & N &  0.924 & 5.27 & -0.629 & 3.75 & 2.81 & 3.65 & 2.35 & 4.03 &  0.133 & 2.32 & 2.12 &  0.977 & 1.42 \\
117176 & N &  0.968 & 5.89 &  -1.26 & 3.62 & 2.73 & 3.40 & 1.85 & 3.74 &  0.111 & 2.38 & 1.70 &  0.708 & 1.00 \\
117936 & N & 3.85 & 6.30 &  -1.26 & 9.57 & 5.22 & -0.730 & 1.51 & 5.11 &  0.471 & 8.22 & 4.74 & 3.38 & 5.85 \\
118914 & N &  0.816 & 4.84 & -0.211 & 3.72 & 2.77 &  0.454 & 2.71 & 4.14 &  0.152 & 2.25 & 2.56 & 1.19 & 1.55 \\
120066 & N &  0.873 & 4.54 &  0.517 & 2.88 & 2.91 & 2.81 & 2.56 & 4.37 &  0.101 & 2.13 & 1.71 &  0.724 & 1.00 \\
120136 & N &  0.654 & 2.38 & 3.07 & 1.78 & 2.65 & 1.72 & 4.12 & 4.19 & 0.0760 & 1.21 & 1.50 &  0.429 & 1.07 \\
121560 & N &  0.453 & 3.26 & 1.72 & 1.44 & 1.71 &  0.118 & 3.13 & 2.56 & 0.0749 & 1.49 &  0.875 &  0.592 &  0.932 \\
122120 & N & 4.84 & 4.39 &  -1.02 & 9.02 & 5.26 &  -1.91 &  0.317 & 6.62 &  0.563 & 8.89 & 4.92 & 3.82 & 8.01 \\
122652 & N &  0.465 & 2.70 & 1.81 & 2.41 & 2.25 & 1.54 & 3.52 & 2.56 & 0.0894 & 1.83 & 1.64 &  0.649 & 1.10 \\
122676 & N & 1.14 & 5.60 &  -1.37 & 4.33 & 2.96 & 3.82 & 2.20 & 4.63 &  0.174 & 3.55 & 2.31 & 1.65 & 2.30 \\
124642 & N & 4.88 & 6.09 &  -1.15 & 9.67 & 4.99 &  -1.25 & 1.37 & 5.17 &  0.472 & 7.78 & 5.05 & 3.51 & 5.90 \\
124694 & N &  0.474 & 3.28 & 1.68 & 1.95 & 2.10 & 2.17 & 3.37 & 3.79 & 0.0836 & 1.72 & 1.66 &  0.544 &  0.992 \\
125040 & N &  0.652 & 1.91 & 2.54 & 1.71 & 2.03 &  0.927 & 3.71 & 3.37 & 0.0900 & 1.42 & 1.70 &  0.767 & 1.13 \\
126053 & N & 1.08 & 5.29 & -0.550 & 3.06 & 1.90 & -0.757 & 2.15 & 2.27 &  0.123 & 2.61 & 1.56 &  0.480 & 1.25 \\
126961 & N &  0.780 & 3.23 & 1.98 & 2.32 & 1.99 & 2.16 & 3.48 & 4.24 & 0.0892 & 1.60 & 1.60 &  0.801 & 1.22 \\
127334 & N &  0.793 & 5.56 &  -1.11 & 4.31 & 2.71 & 5.91 & 2.36 & 5.25 &  0.148 & 2.88 & 2.25 & 1.15 & 1.60 \\
128165 & N & 3.86 & 6.42 &  -1.39 & 8.34 & 4.28 &  -1.51 &  0.730 & 4.94 &  0.458 & 8.34 & 4.41 & 3.22 & 5.21 \\
130087 & N &  0.692 & 4.68 &  0.773 & 2.79 & 2.59 & 4.04 & 3.02 & 5.97 &  0.119 & 1.96 & 2.14 &  0.919 & 1.66 \\
130307 & N & 2.97 & 6.79 &  -1.34 & 6.87 & 3.71 &  -1.20 & 1.10 & 3.46 &  0.330 & 7.00 & 3.61 & 2.08 & 3.37 \\
130322 & N & 1.71 & 6.98 &  -1.38 & 6.39 & 3.34 & 4.76 & 1.77 & 3.73 &  0.219 & 4.48 & 3.13 & 1.88 & 2.45 \\
130871 & N & 3.21 & 6.64 &  -1.02 & 7.29 & 3.00 &  -1.29 &  0.666 & 3.59 &  0.441 & 8.67 & 4.27 & 2.60 & 4.47 \\
131509 & N & 1.29 & 7.36 &  -1.75 & 5.58 & 2.99 & 5.37 & 1.43 & 3.12 &  0.217 & 4.12 & 3.12 & 1.77 & 2.16 \\
132142 & N & 1.90 & 6.02 &  -2.49 & 5.64 & 2.91 & 3.58 & 1.27 & 4.19 &  0.287 & 6.24 & 3.03 & 1.63 & 2.39 \\
133161 & N &  0.730 & 4.30 &  0.888 & 2.94 & 2.54 & 3.83 & 3.27 & 4.77 &  0.116 & 2.31 & 2.11 &  0.863 & 1.48 \\
  & K &  0.595 & 4.34 &  0.800 & 2.92 & 2.84 & 3.56 & 3.33 & 4.68 &  0.104 & 1.69 & 1.86 &  0.820 & 1.69 \\
133460 & N &  0.810 & 4.22 & 1.46 & 2.75 & 2.43 & 2.55 & 3.42 & 5.10 &  0.103 & 1.84 & 1.79 &  0.973 & 1.36 \\
134044 & N &  0.589 & 3.20 & 1.94 & 1.93 & 2.12 & 2.27 & 3.56 & 4.31 & 0.0841 & 1.47 & 1.71 &  0.590 & 1.00 \\
135101 & N &  0.799 & 5.52 & -0.952 & 3.79 & 2.60 & 4.92 & 2.27 & 4.78 &  0.150 & 2.96 & 1.94 &  0.901 & 1.29 \\
135599 & N & 2.22 & 6.71 &  -1.71 & 6.07 & 3.23 & -0.668 & 1.51 & 3.88 &  0.254 & 5.15 & 3.15 & 1.81 & 2.38 \\
136118 & K &  0.438 & 3.62 & 1.65 & 2.13 & 2.21 & 1.50 & 3.38 & 3.30 & 0.0633 &  0.883 & 1.25 &  0.401 & 1.15 \\
136442 & K & 2.25 & 6.24 &  -2.15 & 8.44 & 4.43 & 9.95 & 1.16 & 7.13 &  0.363 & 6.83 & 4.55 & 2.93 & 4.44 \\
136544 & N &  0.610 &  0.327 & 4.47 & 1.51 & 2.19 & 1.54 & 4.15 & 4.24 & 0.0787 & 1.46 & 1.65 &  0.541 &  0.994 \\
  & K &  0.356 & 2.61 & 3.29 &  0.954 & 2.36 & 1.51 & 4.15 & 3.96 & 0.0628 & 1.11 & 1.38 &  0.395 & 1.22 \\
136580 & N &  0.271 & 2.57 & 2.11 & 1.67 & 1.74 & 1.07 & 3.48 & 3.54 & 0.0750 & 1.48 & 1.25 &  0.443 &  0.947 \\
136654 & N &  0.538 & 2.70 & 2.54 & 1.39 & 2.13 & 1.96 & 3.97 & 4.14 & 0.0726 & 1.17 & 1.44 &  0.537 & 1.02 \\
136834 & N & 3.27 & 6.93 &  -1.62 & 10.1 & 3.67 & 7.58 & 1.04 & 4.70 &  0.456 & 8.49 & 4.53 & 3.24 & 5.70 \\
136923 & N & 1.87 & 6.52 &  -1.67 & 5.59 & 3.20 & 4.00 & 1.76 & 3.45 &  0.219 & 4.54 & 2.70 & 1.64 & 2.16 \\
137510 & N &  0.113 & 2.90 & 2.53 & 3.35 & 2.52 & 4.37 & 3.37 & 5.73 &  0.125 & 2.30 & 2.15 &  0.994 & 1.43 \\
  & K &  0.435 & 4.84 &  0.604 & 3.04 & 3.13 & 4.69 & 3.25 & 5.29 &  0.108 & 1.62 & 1.99 &  0.855 & 1.62 \\
137778 & K & 2.44 & 5.86 &  -1.55 & 8.10 & 4.21 & 6.46 & 1.43 & 6.59 &  0.315 & 5.48 & 3.93 & 2.63 & 4.67 \\
138573 & N &  0.847 & 5.39 & -0.569 & 4.23 & 2.37 & 3.43 & 2.44 & 3.05 &  0.142 & 2.56 & 2.25 & 1.07 & 1.56 \\
139323 & N & 3.20 & 6.50 &  -3.09 & 8.44 & 4.57 & 8.86 & 1.29 & 8.03 &  0.395 & 6.65 & 4.64 & 3.15 & 4.72 \\
139324 & N &  0.755 & 5.02 &  0.240 & 2.99 & 2.55 & 3.57 & 2.89 & 5.18 &  0.116 & 1.96 & 2.08 & 1.01 & 1.45 \\
  & K &  0.520 & 4.80 &  -0.0381 & 3.86 & 2.60 & 3.43 & 2.98 & 4.55 &  0.113 & 1.93 & 1.88 &  0.781 & 1.55 \\
139457 & N &  0.429 & 3.02 & 1.71 & 1.70 & 1.84 &  0.496 & 2.97 & 1.52 & 0.0706 & 1.36 & 1.21 &  0.523 & 1.05 \\
  & K &  0.320 & 3.49 & 1.68 & 1.34 & 1.78 &  0.464 & 3.15 & 2.54 & 0.0648 & 1.02 & 1.13 &  0.242 &  0.922 \\
142229 & N &  0.951 & 4.46 &  0.391 & 3.33 & 2.04 &  -1.09 & 2.53 & 4.57 &  0.120 & 2.21 & 2.34 & 1.11 & 1.67 \\
  & N &  0.731 & 3.89 &  0.787 & 2.27 & 2.31 & 2.01 & 2.74 & 4.60 &  0.109 & 2.53 & 2.32 &  0.962 & 1.61 \\
  & K &  0.807 & 4.39 & 0.0796 & 3.58 & 2.68 & 2.67 & 2.71 & 4.24 &  0.105 & 1.96 & 1.78 &  0.888 & 1.55 \\
142373 & N &  0.154 & 4.51 &  0.520 & 1.84 & 1.84 &  0.574 & 2.28 & 3.11 & 0.0841 & 1.52 & 1.45 &  0.467 & 1.06 \\
143291 & N & 1.66 & 6.33 &  -1.58 & 5.31 & 2.63 & -0.778 & 1.59 & 4.12 &  0.227 & 4.80 & 2.49 & 1.42 & 1.94 \\
  & N & 1.46 & 6.08 &  -1.20 & 4.43 & 2.43 & 2.93 & 1.26 & 4.06 &  0.226 & 5.64 & 2.61 & 1.52 & 2.03 \\
143761 & N &  0.794 & 4.66 & 0.0369 & 2.62 & 2.14 & 1.82 & 2.41 & 3.74 &  0.108 & 2.27 & 1.38 &  0.399 & 1.05 \\
  & K &  0.660 & 4.76 & -0.191 & 3.09 & 2.25 & 1.79 & 2.44 & 3.59 &  0.101 & 2.11 & 1.37 &  0.637 & 1.19 \\
144579 & N & 1.58 & 5.76 &  -2.31 & 3.97 & 2.05 & 1.86 & 1.38 & 2.20 &  0.208 & 5.12 & 1.77 &  0.893 & 1.69 \\
145229 & N &  0.652 & 3.96 &  0.453 & 2.64 & 2.66 & 1.48 & 2.58 & 3.98 &  0.104 & 2.09 & 1.62 &  0.624 & 1.33 \\
148467 & N & 6.14 & 4.73 &  -1.52 & 8.05 & 5.68 &  -2.99 & 1.79 & 4.73 &  0.581 & 9.23 & 5.04 & 4.04 & 8.80 \\
  & K & 6.91 & 5.55 &  -1.35 & 8.63 & 5.56 &  -3.24 & 1.33 & 5.90 &  0.559 & 8.38 & 5.00 & 3.91 & 9.80 \\
149200 & K &  0.600 & 2.90 & 2.35 & 2.32 & 2.40 & 2.07 & 3.73 & 4.00 & 0.0665 & 1.37 & 1.38 &  0.521 & 1.04 \\
149652 & N &  0.435 & 2.84 & 2.35 & 1.57 & 1.77 &  0.811 & 3.60 & 2.95 & 0.0745 &  0.941 & 1.48 &  0.490 &  0.964 \\
149661 & N & 1.43 & 6.35 &  -1.01 & 5.66 & 3.35 & 4.51 & 1.42 & 5.62 &  0.270 & 5.39 & 3.31 & 2.05 & 2.95 \\
149806 & N & 1.60 & 6.74 &  -1.64 & 6.86 & 3.34 & 6.78 & 1.76 & 4.03 &  0.270 & 5.42 & 3.32 & 2.10 & 3.04 \\
150933 & N &  0.841 & 3.88 & 1.33 & 2.60 & 2.37 & 2.46 & 3.29 & 4.49 &  0.104 & 1.67 & 1.64 &  0.721 & 1.30 \\
151044 & N &  0.709 & 3.75 & 1.64 & 2.16 & 2.02 & 1.72 & 3.15 & 3.86 & 0.0922 & 1.47 & 1.59 &  0.708 & 1.01 \\
151090 & N & 1.40 & 7.60 &  -1.85 & 5.66 & 2.76 & 4.18 & 1.22 & 4.67 &  0.200 & 3.82 & 2.36 & 1.28 & 1.51 \\
  & K & 1.25 & 6.02 &  -1.86 & 5.27 & 3.21 & 4.44 & 1.31 & 4.70 &  0.203 & 3.74 & 2.62 & 1.51 & 1.93 \\
  & K & 1.22 & 6.16 &  -1.62 & 5.21 & 3.19 & 4.76 & 1.30 & 4.40 &  0.202 & 3.90 & 2.74 & 1.51 & 2.01 \\
151288 & N & 7.44 & 5.26 &  -1.32 & 8.42 & 5.26 &  -2.21 & 1.25 & 6.50 &  0.552 & 7.86 & 5.07 & 4.29 & 10.8 \\
151877 & N & 2.31 & 6.98 &  -1.78 & 5.78 & 2.84 & 4.28 & 1.61 & 5.78 &  0.264 & 5.39 & 3.29 & 1.70 & 2.73 \\
  & K & 1.96 & 5.58 &  -1.54 & 6.58 & 3.39 & 3.90 & 1.46 & 5.11 &  0.250 & 5.34 & 3.36 & 1.97 & 3.06 \\
152446 & N &  0.575 & 2.68 & 1.96 & 1.80 & 1.80 & 1.49 & 3.40 & 3.68 & 0.0746 & 1.17 & 1.19 &  0.355 &  0.936 \\
  & N &  0.508 & 2.77 & 1.96 & 1.12 & 1.79 &  0.800 & 3.30 & 3.71 & 0.0851 & 1.15 & 1.37 &  0.445 & 1.02 \\
152792 & K &  0.500 & 5.06 & -0.609 & 3.29 & 2.19 & 1.54 & 2.21 & 3.38 & 0.0910 & 1.77 & 1.48 &  0.587 & 1.17 \\
153627 & N &  0.612 & 4.29 &  0.859 & 2.63 & 1.75 &  0.733 & 2.91 & 3.66 & 0.0985 & 1.56 & 1.59 &  0.496 &  0.975 \\
154160 & N &  0.861 & 5.62 &  -1.48 & 4.78 & 3.02 & 7.60 & 2.49 & 3.82 &  0.166 & 2.69 & 2.49 & 1.24 & 1.77 \\
  & N &  0.546 & 5.75 &  -1.16 & 4.20 & 2.76 & 7.51 & 2.48 & 6.46 &  0.189 & 3.35 & 2.92 & 1.53 & 2.14 \\
154345 & N & 1.25 & 5.50 &  -1.56 & 4.24 & 2.64 & 3.24 & 1.93 & 4.84 &  0.196 & 3.97 & 2.48 & 1.43 & 2.04 \\
154363 & K & 6.56 & 5.56 &  -1.72 & 7.37 & 4.63 &  -2.61 & 1.09 & 5.84 &  0.592 & 9.59 & 4.50 & 3.19 & 8.28 \\
  & K & 6.34 & 5.12 &  -1.56 & 7.34 & 4.60 &  -2.32 & 1.04 & 5.79 &  0.600 & 9.84 & 4.44 & 3.12 & 8.25 \\
154417 & N &  0.850 & 3.90 & 1.02 & 2.98 & 2.27 & 1.73 & 3.12 & 4.67 &  0.101 & 1.87 & 1.92 &  0.844 & 1.08 \\
155060 & N &  0.843 & 3.99 & 1.01 & 2.18 & 2.07 & 1.37 & 2.98 & 2.76 & 0.0942 & 1.67 & 1.55 &  0.637 & 1.06 \\
155423 & N &  0.666 & 3.37 & 1.97 & 2.38 & 2.23 & 2.92 & 3.66 & 5.10 & 0.0988 & 1.53 & 1.76 &  0.503 & 1.17 \\
156826 & K &  0.889 & 6.10 &  -1.82 & 5.24 & 3.04 & 3.78 & 1.41 & 4.41 &  0.157 & 3.30 & 2.61 & 1.35 & 2.06 \\
157214 & N & 1.06 & 5.19 & -0.397 & 2.93 & 2.14 & 2.31 & 2.35 & 3.95 &  0.126 & 2.80 & 1.42 &  0.555 & 1.08 \\
157466 & N &  0.569 & 3.14 & 1.54 & 1.63 & 1.58 &  0.757 & 2.94 & 1.87 & 0.0795 & 1.41 &  0.897 &  0.384 &  0.923 \\
157881 & N & 6.62 & 4.62 &  -1.48 & 7.86 & 5.57 &  -3.13 & 1.13 & 5.54 &  0.545 & 7.68 & 5.01 & 3.80 & 10.7 \\
  & K & 6.36 & 4.86 &  -1.20 & 7.67 & 5.50 &  -3.11 & 1.30 & 5.79 &  0.529 & 7.18 & 5.14 & 4.00 & 12.3 \\
  & K & 6.33 & 4.73 &  -1.15 & 7.05 & 5.53 &  -3.08 & 1.28 & 5.81 &  0.526 & 7.15 & 5.24 & 4.03 & 12.5 \\
159063 & N &  0.642 & 3.14 & 2.39 & 2.04 & 2.28 & 2.43 & 3.85 & 4.98 & 0.0943 & 1.48 & 1.63 &  0.660 & 1.15 \\
159222 & N &  0.952 & 4.78 & 0.0173 & 4.02 & 2.33 & 3.39 & 2.76 & 4.77 &  0.123 & 2.04 & 1.95 &  0.816 & 1.38 \\
159909 & N & 1.11 & 5.46 & -0.780 & 3.87 & 2.80 & 4.73 & 2.65 & 5.26 &  0.155 & 2.85 & 2.16 & 1.09 & 1.64 \\
160693 & N &  0.720 & 5.07 &  -0.0442 & 2.57 & 1.71 & 1.08 & 2.50 & 3.80 &  0.106 & 1.92 & 1.05 &  0.316 &  0.899 \\
161797 & N & 1.03 & 5.20 &  -1.12 & 4.49 & 2.90 & 7.12 & 2.48 & 5.98 &  0.174 & 2.77 & 2.75 & 1.49 & 1.90 \\
161848 & N & 2.42 & 6.78 &  -1.76 & 5.88 & 3.10 & 2.47 & 1.27 & 4.24 &  0.314 & 6.52 & 3.28 & 1.76 & 2.51 \\
162826 & N &  0.786 & 3.36 & 1.68 & 1.79 & 1.99 & 1.83 & 3.14 & 3.78 & 0.0809 & 1.26 & 1.31 &  0.539 &  0.883 \\
164507 & K &  0.694 & 5.66 &  -1.45 & 4.58 & 3.02 & 4.94 & 2.26 & 4.97 &  0.126 & 2.25 & 2.31 & 1.12 & 1.68 \\
164922 & N & 1.96 & 5.72 &  -1.89 & 6.33 & 3.26 & 5.38 & 1.79 & 4.00 &  0.245 & 4.62 & 3.22 & 1.85 & 2.31 \\
165567 & N &  0.473 & 2.45 & 2.72 & 1.43 & 1.90 & 1.47 & 3.61 & 3.38 & 0.0681 &  0.918 & 1.12 &  0.245 &  0.756 \\
  & K &  0.336 & 2.60 & 2.52 & 1.81 & 2.25 & 1.33 & 3.83 & 3.69 & 0.0607 &  0.851 & 1.19 &  0.421 & 1.03 \\
166435 & N & 1.13 & 4.64 &  -0.0214 & 3.20 & 2.57 & 2.46 & 2.56 & 4.63 &  0.126 & 2.17 & 2.16 & 1.03 & 1.53 \\
167215 & N &  0.395 & 2.99 & 2.13 & 1.10 & 1.49 & -0.119 & 3.31 & 3.25 & 0.0707 &  0.933 & 1.18 &  0.711 & 1.07 \\
167389 & N & 1.06 & 4.61 &  0.420 & 3.02 & 2.34 & 2.57 & 2.87 & 4.32 &  0.119 & 2.23 & 1.90 &  0.737 & 1.20 \\
169822 & N & 1.23 & 5.87 &  -1.24 & 4.26 & 2.38 & 3.02 & 1.92 & 3.52 &  0.170 & 3.28 & 2.42 & 1.35 & 1.85 \\
  & N &  0.869 & 5.36 &  -1.21 & 3.82 & 2.15 & 3.01 & 2.13 & 3.97 &  0.166 & 3.66 & 2.29 & 1.03 & 1.93 \\
170469 & N &  0.839 & 5.58 & -0.639 & 4.03 & 3.11 & 5.15 & 2.87 & 3.05 &  0.160 & 2.65 & 2.53 & 1.07 & 1.90 \\
  & K &  0.660 & 5.17 & -0.642 & 4.66 & 3.01 & 6.03 & 2.80 & 5.24 &  0.150 & 2.92 & 2.38 & 1.17 & 2.15 \\
170778 & K &  0.819 & 3.94 &  0.209 & 3.62 & 2.56 & 1.93 & 2.80 & 3.78 & 0.0992 & 1.79 & 1.72 &  0.800 & 1.37 \\
170829 & N & 1.08 & 6.20 &  -2.05 & 5.18 & 2.81 & 5.78 & 1.98 & 4.74 &  0.158 & 2.70 & 2.30 & 1.12 & 1.46 \\
  & K &  0.810 & 5.92 &  -2.53 & 5.61 & 3.07 & 5.98 & 2.00 & 5.07 &  0.167 & 2.86 & 2.74 & 1.40 & 2.05 \\
171067 & N & 1.35 & 5.11 & -0.869 & 4.24 & 2.57 & 3.00 & 2.38 & 4.53 &  0.156 & 3.05 & 2.42 & 1.13 & 1.63 \\
171918 & K &  0.681 & 5.69 &  -1.20 & 4.84 & 3.12 & 4.91 & 2.70 & 4.92 &  0.139 & 2.19 & 2.20 & 1.09 & 1.84 \\
172310 & N & 1.31 & 5.75 &  -2.21 & 4.57 & 2.19 & 2.26 & 1.58 & 2.98 &  0.178 & 4.32 & 2.51 &  0.998 & 1.93 \\
173701 & N & 1.53 & 7.19 &  -1.76 & 5.94 & 3.69 & 8.35 & 2.05 & 3.75 &  0.281 & 5.07 & 3.53 & 2.07 & 3.41 \\
173818 & K & 6.73 & 4.64 &  -1.18 & 7.60 & 5.47 &  -3.02 & 1.13 & 5.05 &  0.531 & 7.90 & 4.75 & 3.58 & 9.74 \\
174080 & N & 4.64 & 6.77 &  -1.76 & 9.49 & 4.94 & -0.862 & 1.76 & 5.01 &  0.502 & 8.53 & 5.08 & 3.82 & 6.56 \\
  & K & 5.08 & 5.38 &  -1.60 & 9.32 & 5.03 &  -1.85 & 1.68 & 6.41 &  0.483 & 8.34 & 4.98 & 3.68 & 7.63 \\
174457 & N &  0.799 & 4.45 &  0.574 & 2.69 & 1.98 & 1.49 & 2.63 & 3.90 &  0.111 & 1.88 & 1.55 &  0.714 & 1.27 \\
174912 & N &  0.515 & 3.49 &  0.731 & 2.20 & 1.77 &  0.810 & 2.64 & 2.94 & 0.0899 & 1.66 & 1.24 &  0.561 & 1.05 \\
175317 & K &  0.250 & 2.03 & 3.62 & 1.17 & 2.10 &  0.820 & 4.36 & 3.34 & 0.0453 &  0.590 & 1.08 &  0.322 & 1.23 \\
175541 & K &  0.878 & 6.42 &  -1.76 & 5.53 & 3.36 & 4.51 & 1.44 & 5.25 &  0.162 & 2.65 & 2.71 & 1.57 & 1.97 \\
175726 & N &  0.879 & 3.52 & 1.13 & 2.46 & 2.05 & 1.30 & 2.75 & 4.04 & 0.0930 & 1.66 & 1.44 &  0.549 & 1.04 \\
176377 & N &  0.938 & 4.23 &  0.262 & 2.73 & 1.63 & 1.24 & 2.56 & 2.16 &  0.109 & 2.05 & 1.41 &  0.562 & 1.23 \\
177830 & K & 1.78 & 6.32 &  -1.92 & 8.60 & 4.36 & 9.67 & 1.39 & 7.29 &  0.343 & 5.79 & 4.29 & 2.94 & 4.65 \\
181655 & K & 1.03 & 4.80 &  -1.02 & 4.29 & 2.84 & 3.82 & 2.47 & 4.39 &  0.143 & 3.16 & 2.47 & 1.16 & 1.54 \\
182488 & K & 1.41 & 5.57 &  -1.70 & 5.27 & 3.33 & 6.27 & 1.96 & 5.65 &  0.240 & 5.06 & 3.39 & 1.83 & 3.04 \\
183650 & N & 0.000900 & 5.23 & -0.752 & 3.66 & 2.60 & 6.27 & 2.54 & 6.14 &  0.183 & 2.95 & 2.96 & 1.33 & 2.04 \\
  & N &  0.537 & 2.54 & 2.15 & 4.23 & 3.19 & 6.44 & 2.47 & 5.87 &  0.183 & 3.70 & 2.77 & 1.27 & 2.03 \\
  & K &  0.527 & 5.68 &  -1.32 & 4.72 & 3.12 & 6.94 & 2.69 & 5.15 &  0.171 & 2.95 & 2.92 & 1.20 & 2.18 \\
183658 & N &  0.691 & 4.89 & -0.343 & 3.31 & 2.34 & 3.42 & 2.61 & 4.92 &  0.145 & 2.78 & 2.22 &  0.953 & 1.57 \\
184385 & N &  0.918 & 3.35 & 1.58 & 4.33 & 2.72 & 4.25 & 2.11 & 5.20 &  0.196 & 3.54 & 2.95 & 1.46 & 2.08 \\
184860 & N & 2.09 & 3.76 & 2.51 & 6.87 & 3.64 & 3.58 &  0.609 & 5.09 &  0.445 & 7.97 & 4.37 & 2.86 & 4.77 \\
188512 & K &  0.981 & 6.39 &  -1.59 & 5.77 & 3.20 & 4.46 & 1.52 & 4.81 &  0.174 & 3.17 & 2.67 & 1.40 & 1.98 \\
189067 & K &  0.642 & 5.04 &  0.156 & 2.45 & 2.55 & 3.01 & 2.90 & 3.86 &  0.105 & 1.65 & 1.68 &  0.762 & 1.03 \\
190007 & K & 5.61 & 5.48 &  -1.36 & 8.93 & 5.49 &  -1.99 & 1.56 & 6.43 &  0.524 & 8.17 & 5.07 & 3.92 & 8.79 \\
190067 & N & 1.48 & 5.62 &  -1.71 & 4.46 & 2.72 & 2.65 & 1.75 & 4.35 &  0.195 & 4.39 & 2.48 & 1.33 & 1.98 \\
190228 & N &  0.853 & 6.28 &  -1.93 & 4.35 & 2.37 & 3.61 & 1.50 & 4.31 &  0.157 & 2.72 & 2.44 & 1.26 & 1.59 \\
  & K &  0.837 & 6.26 &  -2.68 & 4.92 & 2.94 & 3.38 & 1.69 & 4.37 &  0.145 & 2.78 & 2.28 & 1.15 & 1.64 \\
190360 & K &  0.822 & 5.37 &  -1.90 & 5.37 & 3.31 & 6.50 & 2.36 & 5.39 &  0.196 & 3.46 & 2.79 & 1.41 & 2.06 \\
190771 & K & 1.01 & 4.50 & -0.471 & 4.13 & 2.97 & 3.73 & 2.65 & 4.60 &  0.133 & 2.61 & 2.35 & 1.14 & 1.47 \\
191022 & K &  0.597 & 5.05 & -0.120 & 2.82 & 2.80 & 3.44 & 2.79 & 4.45 &  0.116 & 2.06 & 2.14 &  0.856 & 1.58 \\
192343 & K &  0.675 & 5.37 & -0.533 & 3.44 & 2.89 & 5.10 & 2.87 & 4.95 &  0.138 & 2.17 & 2.34 & 1.01 & 1.67 \\
194035 & N &  0.908 & 5.93 & -0.953 & 4.15 & 2.78 & 5.81 & 2.36 & 5.43 &  0.148 & 2.53 & 2.60 & 1.28 & 1.49 \\
195019 & N &  0.928 & 4.85 & -0.288 & 3.24 & 2.56 & 3.06 & 2.36 & 4.76 &  0.125 & 2.17 & 2.13 & 1.01 & 1.32 \\
195104 & N &  0.730 & 2.65 & 1.95 & 1.78 & 2.03 & 1.53 & 3.31 & 3.74 & 0.0813 & 1.40 & 1.27 &  0.579 & 1.08 \\
196201 & N & 1.17 & 5.61 &  -1.47 & 4.26 & 3.07 & 3.14 & 1.66 & 4.13 &  0.192 & 3.80 & 2.61 & 1.53 & 2.25 \\
196850 & N &  0.874 & 4.28 &  0.110 & 3.09 & 2.40 & 2.59 & 2.47 & 4.13 &  0.112 & 2.21 & 1.79 &  0.789 & 1.27 \\
196885 & N &  0.823 & 3.07 & 2.02 & 2.04 & 2.44 & 2.54 & 3.33 & 3.91 & 0.0786 & 1.35 & 1.42 &  0.471 & 1.04 \\
197076 & N &  0.965 & 4.56 &  0.141 & 2.99 & 2.19 & 2.21 & 2.40 & 3.72 &  0.106 & 2.07 & 1.60 &  0.589 & 1.16 \\
198089 & N &  0.718 & 4.31 &  0.172 & 2.47 & 2.17 & 1.67 & 2.87 & 3.80 &  0.107 & 2.08 & 1.69 &  0.796 & 1.37 \\
198387 & K & 1.15 & 6.33 &  -1.61 & 6.13 & 3.33 & 5.05 & 1.33 & 5.14 &  0.198 & 3.68 & 2.85 & 1.67 & 2.00 \\
198802 & K &  0.426 & 5.59 & -0.654 & 3.69 & 2.71 & 3.13 & 2.59 & 4.22 &  0.102 & 1.49 & 2.05 &  0.769 & 1.44 \\
199598 & N &  0.874 & 3.95 &  0.979 & 3.30 & 2.39 & 3.16 & 2.91 & 4.51 &  0.110 & 1.90 & 1.80 &  0.908 & 1.30 \\
202108 & K &  0.940 & 4.93 & -0.920 & 3.75 & 2.54 & 2.07 & 2.28 & 3.60 &  0.123 & 2.34 & 1.99 &  0.865 & 1.29 \\
202575 & K & 4.72 & 5.30 &  -1.33 & 8.80 & 4.76 & 3.28 & 1.48 & 5.57 &  0.433 & 7.73 & 4.55 & 3.27 & 5.79 \\
204587 & K & 6.89 & 5.32 &  -1.34 & 8.12 & 5.74 &  -2.82 &  0.891 & 5.98 &  0.579 & 8.27 & 5.17 & 3.82 & 10.3 \\
  & K & 6.94 & 5.13 &  -1.27 & 7.19 & 5.65 &  -2.75 &  0.930 & 6.10 &  0.576 & 8.26 & 5.16 & 3.81 & 10.4 \\
207740 & K & 1.17 & 5.58 &  -2.01 & 5.20 & 2.99 & 4.01 & 2.08 & 4.63 &  0.166 & 2.97 & 2.27 & 1.30 & 1.99 \\
  & K & 1.13 & 5.64 &  -1.90 & 5.32 & 2.90 & 4.07 & 2.09 & 4.69 &  0.169 & 2.87 & 2.54 & 1.33 & 2.19 \\
208776 & K &  0.325 & 4.54 &  0.603 & 2.45 & 2.38 & 2.13 & 3.02 & 3.62 & 0.0824 & 1.31 & 1.60 &  0.608 & 1.09 \\
208801 & K & 1.67 & 6.06 &  -1.83 & 6.93 & 3.86 & 7.78 & 1.23 & 5.80 &  0.314 & 5.38 & 3.65 & 2.38 & 3.26 \\
209393 & K & 1.08 & 5.18 &  -1.42 & 4.58 & 2.78 & 2.52 & 2.16 & 4.11 &  0.141 & 2.43 & 1.92 & 1.06 & 1.59 \\
210312 & K &  0.813 & 5.32 &  -1.16 & 5.29 & 3.23 & 5.28 & 2.73 & 5.35 &  0.154 & 2.63 & 2.36 & 1.22 & 2.13 \\
210460 & K &  0.576 & 5.34 &  -1.29 & 3.57 & 2.67 & 2.26 & 1.96 & 3.80 &  0.100 & 1.78 & 1.78 &  0.844 & 1.29 \\
210667 & K & 1.49 & 6.03 &  -1.23 & 6.68 & 3.74 & 5.95 & 1.90 & 5.96 &  0.235 & 4.66 & 3.30 & 1.90 & 2.88 \\
211038 & K & 1.46 & 6.06 &  -1.78 & 5.45 & 3.33 & 4.57 & 1.32 & 4.62 &  0.246 & 5.19 & 3.04 & 1.67 & 2.42 \\
211080 & K &  0.520 & 5.54 & -0.751 & 4.13 & 3.14 & 6.11 & 2.72 & 5.23 &  0.142 & 2.27 & 2.45 & 1.17 & 1.96 \\
  & K &  0.584 & 5.32 &  -1.09 & 4.89 & 3.21 & 5.72 & 2.70 & 5.68 &  0.143 & 2.12 & 2.54 & 1.07 & 2.19 \\
213472 & K &  0.844 & 5.54 & -0.846 & 4.37 & 2.95 & 3.96 & 2.56 & 4.88 &  0.129 & 1.95 & 1.98 &  0.980 & 1.65 \\
  & K &  0.902 & 5.33 & -0.943 & 4.54 & 2.87 & 3.95 & 2.63 & 4.63 &  0.128 & 2.41 & 2.17 & 1.04 & 1.41 \\
216259 & K & 2.82 & 5.59 &  -1.41 & 5.23 & 3.06 & 1.72 &  0.749 & 3.74 &  0.343 & 6.84 & 2.86 & 1.63 & 3.09 \\
217014 & K &  0.707 & 4.94 & -0.647 & 4.32 & 2.93 & 4.66 & 2.81 & 4.47 &  0.143 & 2.33 & 2.52 & 1.07 & 1.80 \\
218687 & N &  0.690 & 3.74 &  0.395 & 1.87 & 2.25 & 1.87 & 2.94 & 4.02 &  0.101 & 1.94 & 1.82 &  0.998 & 1.42 \\
  & N &  0.690 & 3.74 &  0.395 & 1.87 & 2.25 & 1.87 & 2.94 & 4.02 &  0.101 & 1.94 & 1.82 &  0.998 & 1.42 \\
219172 & N &  0.460 & 2.99 & 1.52 & 1.68 & 1.99 & 2.66 & 3.51 & 4.41 & 0.0910 & 1.46 & 1.62 &  0.745 & 1.21 \\
  & N &  0.460 & 2.99 & 1.52 & 1.68 & 1.99 & 2.66 & 3.51 & 4.41 & 0.0910 & 1.46 & 1.62 &  0.745 & 1.21 \\
220339 & N & 1.66 & 4.47 & 1.66 & 6.57 & 3.23 &  -1.41 & 1.04 & 4.69 &  0.346 & 7.48 & 3.53 & 2.25 & 3.47 \\
  & N & 1.66 & 4.47 & 1.66 & 6.57 & 3.23 &  -1.41 & 1.04 & 4.69 &  0.346 & 7.48 & 3.53 & 2.25 & 3.47 \\
221146 & K &  0.571 & 4.91 & -0.266 & 3.97 & 2.82 & 3.20 & 2.84 & 4.25 &  0.117 & 2.12 & 1.96 &  0.848 & 1.68 \\
221830 & N &  0.541 & 4.64 &  0.192 & 1.91 & 1.79 & 1.22 & 2.25 & 3.70 &  0.133 & 2.79 & 1.52 & 1.06 & 1.06 \\
  & N &  0.550 & 4.03 &  0.150 & 1.40 & 1.87 & 1.38 & 2.33 & 3.47 &  0.121 & 2.87 & 1.37 &  0.565 & 1.07 \\
  & N &  0.550 & 4.03 &  0.150 & 1.40 & 1.87 & 1.38 & 2.33 & 3.47 &  0.121 & 2.87 & 1.37 &  0.565 & 1.07 \\
223084 & N &  0.597 & 2.71 & 1.21 & 1.32 & 1.84 & 1.08 & 3.01 & 3.75 & 0.0813 & 1.74 & 2.63 & 10.4 & 1.34 \\
  & N &  0.597 & 2.71 & 1.21 & 1.32 & 1.84 & 1.08 & 3.01 & 3.75 & 0.0813 & 1.74 & 2.63 & 10.4 & 1.34 \\
223498 & K & 1.05 & 5.51 &  -1.97 & 5.50 & 3.23 & 5.68 & 2.32 & 5.18 &  0.187 & 3.10 & 2.90 & 1.46 & 2.13 \\
225261 & N & 1.25 & 3.94 & 1.39 & 3.30 & 2.46 & 1.79 & 1.37 & 3.99 &  0.212 & 5.45 & 2.59 & 1.28 & 1.92 \\
  & N & 1.25 & 3.94 & 1.39 & 3.30 & 2.46 & 1.79 & 1.37 & 3.99 &  0.212 & 5.45 & 2.59 & 1.28 & 1.92 \\
230409 & K & 1.15 & 5.43 &  -2.11 & 3.58 & 1.92 &  0.576 & 1.49 & 2.62 &  0.165 & 3.90 & 1.53 &  0.747 & 1.58 \\
233641 & N &  0.518 & 3.84 & 1.58 & 1.84 & 2.32 & 1.63 & 3.18 & 3.74 & 0.0753 &  0.934 & 1.64 &  0.372 & 1.13 \\
281540 & K & 4.82 & 5.08 &  -1.16 & 6.22 & 3.58 &  -1.80 & -0.188 & 4.26 &  0.502 & 9.37 & 3.38 & 2.19 & 4.74 \\
\enddata
\tablenotetext{a}{N denotes stars observed with the Nickel telescope at
Lick Observatory, K denotes stars observed with the 2.1m telescope at
Kitt Peak National Observatory}
\tablenotetext{b}{Blank space in HD-number column indicates repeat
observation of star from previous line}
\end{deluxetable}

\clearpage
\begin{table}
\begin{center}
\caption{Polynomial Coefficients and Error Analysis for Atmospheric
Parameter Calibrations.
\label{calibrations}}
\begin{tabular}{lr@{.}lr@{.}lr@{.}l}
\tableline \tableline
Term & \multicolumn{6}{c}{Coefficients} \\
\tableline
& \multicolumn{2}{c}{$T_{\rm eff}$} & \multicolumn{2}{c}{[Fe/H]} &
  \multicolumn{2}{c}{$\log \, g$} \\
\tableline
Ca4227 & -14&3623 & -0&0293059 & 0&194890 \\
G4300 & -14&8610 & 0&0463019 & -0&0789837 \\
H$\gamma_F$ & 18&7798 & 0&0357787 & 0&585302 \\
Fe4383 & 5&47675 & 0&0411910 & -0&00129767 \\
Fe4531 & -55&5761 & 0&130995 & -0&0627585 \\
Fe4668 & 3&23671 & 0&0219853 & -0&00592407 \\
H$\beta$ & 325&223 & 0&0620154 & 0&627797 \\
Fe5015 & 23&2768 & 0&0469375 & -0&0305571 \\
Mg$_2$ & -921&616 & 0&104847 & -1&65883 \\
Mg $b$ & 31&2165 & -0&0710682 & 0&205974 \\
Fe5270 & -3&47323 & 0&182198 & -0&0301877 \\
Na D & -57&6065 & 0&0367736 & 0&0789505 \\
$T_{\rm eff}$ & \multicolumn{2}{c}{ } & 0&000489432 & 0&00112577 \\
$T_{\rm eff} \rm \left ( H\gamma_F + H\beta \right )$ &
  \multicolumn{2}{c}{ } & \multicolumn{2}{c}{ } & -0&000120824 \\
Constant & 5167&60 & -4&21179 & -1&79746 \\
& \multicolumn{2}{c}{ } & \multicolumn{2}{c}{ } & \multicolumn{2}{c}{ }
  \\
\tableline
Reduced $\chi^2$ & 4&52 & 6&75 & 4&10 \\
Gaussian $\sigma$\tablenotemark{a} & \multicolumn{2}{c}{82K} &
  0&07 dex & 0&13 dex \\
Gaussian Center\tablenotemark{a} & -28&4K & -0&017 dex & -0&038 dex \\
Usable Range & \multicolumn{2}{c}{4100--6400K} &
  \multicolumn{2}{c}{-0.95--0.5 dex} &
  \multicolumn{2}{c}{4.8--5.1 dex} \\
\tableline
\end{tabular}
\tablenotetext{a}{From Gaussian fit to histogram of combined test-set
residuals from modified two-phase cross validation method.}
\end{center}
\end{table}

\end{document}